\documentclass[11pt]{article}
\usepackage[letterpaper, left=1in, top=1in, right=1in, bottom=1in,nohead,includefoot]{geometry}
\usepackage{color,charter,graphicx,natbib,here,setspace,multirow,amsmath,amssymb,amsfonts,dsfont,placeins}
\usepackage[usenames,dvipsnames,x11names]{xcolor}
\usepackage[pdftex,pagebackref=true]{hyperref}
\hypersetup{colorlinks,linkcolor=RoyalBlue1,urlcolor=RoyalBlue2, citecolor=RoyalBlue3}

\def\cD{{\cal D}}\def\cI{{\cal I}} \def\cM{{\cal M}} \def\seq#1#2{#1{:}#2}  \def\eqno#1{eqn.~(\ref{eq:#1})}
\def\ind#1{\mathds{1}(#1)} 
\def\F{\mathbf{F}} \def\G{\mathbf{G}} \def\W{\mathbf{W}}  \def\f{\mathbf{f}}
 
\def\I{\mathbf{I}} \def\bzero{\mathbf{0}} 
\def\m{\mathbf{m}} \def\C{\mathbf{C}}\def\a{\mathbf{a}} \def\R{\mathbf{R}} \def\H{\mathbf{H}} \def\P{\mathbf{P}}
\def\btheta{\boldsymbol{\theta}}\def\bomega{\boldsymbol{\omega}} \def\bbeta{\boldsymbol{\beta}} \def\bgamma{\boldsymbol{\gamma}}
\def\bphi{\boldsymbol{\phi}}
\def\bxi{\boldsymbol{\xi}}
\def\ph{\phantom{0}}

\begin{document}

\title{Bayesian forecasting of many count-valued time series}  
 
\author{Lindsay Berry \& Mike West} 
 
\maketitle\thispagestyle{empty}\setcounter{page}0

\begin{abstract}
This paper develops forecasting methodology and application of new classes of dynamic models for time series of non-negative counts. 
Novel univariate models synthesise dynamic generalized linear models for binary and conditionally Poisson time series, with dynamic random effects for over-dispersion. These models allow use of dynamic covariates in both binary and non-zero count components. 
Sequential Bayesian analysis allows fast, parallel analysis of sets of decoupled time series.  New multivariate models then enable information sharing in contexts when data at a more highly aggregated level provide more incisive inferences on shared patterns
such as trends and seasonality. A novel multi-scale approach-- one new example of 
the concept of decouple/recouple in time series-- enables information sharing across series. This incorporates cross-series linkages while insulating  parallel estimation of univariate models, hence enables scalability in the number of series. 
The major motivating context is supermarket sales forecasting. Detailed examples drawn from a case study in multi-step forecasting of sales of a number of related items showcase forecasting of multiple series, with discussion of forecast accuracy metrics and broader questions of probabilistic forecast accuracy assessment.  
\end{abstract}

\noindent%

\vfill
 
\hrule 

\medskip

\footnotesize

\noindent Lindsay Berry (corresponding author) is a PhD candidate and Mike West is 
		The Arts  \& Sciences Professor  of Statistics \& Decision Sciences in the Department of Statistical Science, Duke University, Durham, NC 27708.
		 
		The research reported here was partly supported by $84.51^\circ$, 100 West 5th Street, Cincinnati, OH 45202. The authors acknowledge the input 
		and vision of Dr.~Paul Helman, Chief Science Officer at $84.51^\circ$, and contributions of data sets provided by $84.51^o.$ 
		 
		 Andrew Cron and Natalia Connolly provided useful input on early stages of the   R\&D reported here.  
		 
		 {\em Email:} \href{mailto:Lindsay.Berry@duke.edu}{Lindsay.Berry@duke.edu},
						         \href{mailto:Mike.West@duke.edu}{Mike.West@duke.edu}
\normalsize

\newpage
\section{Introduction}\label{sec:intro}

Modeling and forecasting of multivariate time series of non-negative counts are common interests among many companies and research groups. One key area that motivates our work is that of product sales/demand forecasting, exemplified by forecasting sales based on historical data and concomitant information in commercial outlets including supermarkets and e-commerce sites.   Forecasts for inventory management, production planning, and marketing decisions are at the heart of business analytics in such environments. For large retailers this is a high-dimensional problem as forecasts are required for multiple time granularities for many individual products across multiple outlets. To be effective in such settings, models must run efficiently in an on-line manner as new data is collected, and do so automatically as a routine while having the ability to flag exceptions and call for intervention as needed.  The challenge is to define a flexible class of product-level models that can be customized to individual products within a general framework. Then, forward/sequential analysis and multi-step ahead forecasting must be effective and efficient computationally, and enable integration across potentially many products to share information while maintaining scalability to increasingly large-scale problems.  

Our research to address these challenges begins with definition and development of a novel class of univariate models for time series of non-negative counts. 
Anchored in our case-study context of forecasting daily sales of products at a large supermarket chain, key questions include accounting for various levels of seasonality (weekly, monthly, yearly), holiday effects, price/promotion information, and unpredictable drifts in levels and variability of sales.  High-frequency time series like daily sales are often characterized by high variability and extreme values, and levels of demand across products can vary drastically, with some products selling dozens of units per day, and others having many days with zero sales.   Time series at the fine-scale resolution of individual item sales typically contain many zeros and low counts, so that traditional time series models and methods-- such as exponential smoothing~\citep{hyndman2008forecasting}, ARIMA models~\citep{box2008book}, and conditionally Gaussian/linear state-space models~\citep{west1997book,Prado2010}-- are not appropriate. A number of approaches to forecasting count-valued time series have, of course, been developed. The issues of  intermittent demand (many zeros in sales) and low counts have been a main concern~\citep[e.g.][]{croston1972forecasting}, 
as has over-dispersion relative to Poisson structures. A range of modified Poisson, negative binomial, so-called \lq\lq hurdle shifted'' Poisson and jump-process models have been explored with specific applications~\citep[e.g.][]{chen2016,chen2017,snyder2012forecasting,mccabe2005bayesian}.  Comparative analyses in~\cite{yelland2009bayesian} highlight the utility of state-space approaches including the canonical Gamma-Poisson \lq\lq local level'' model~(\citealp{west1997book} section~10.8; \citealp{Prado2010} section~4.3.7). This and other standard Bayesian state-space models have proven utility in a range of discrete-time series contexts including dynamic network studies~\citep[e.g.][]{ChenETALdynets2016JASA} where short-term forecasting and local smoothing are primary goals. With a view towards improved predictive ability and-- critically-- multi-step forecasting, 
it is perhaps somewhat surprising that more elaborate and predictive  Bayesian state-space models have not yet become central to the area, especially in the context of some of the key genesis developments in Bayesian forecasting in commercial settings~\citep[e.g.][chapter 1, and references therein]{west1997book} and their exploitation over several decades.  

Section~\ref{sec:model} defines and develops the new class of dynamic count mixture models (DCMMs), coupling Bayesian state-space models for binary time series with conditional count models.  We build on standard univariate dynamic generalized linear models~(DGLMs: \citealp{west1985dynamic};~\citealp{west1997book}~chapter~15;~\citealp{Prado2010}~section~4.4) to define a general and flexible class of basic dynamic models that are customizable to individual series.  A critical aspect of DCMMs is that they inherit the sequential learning and forecasting of Bayesian state-space models, allowing fast, parallel processing of decoupled univariate models for individual series.  These models allow for the incorporation of series-specific predictors for zero-count prediction as well as for forecasting levels of non-zero counts,  and-- critically for many applications-- dynamic random effects extensions for over-dispersion relative to conditionally Poisson models.   

The motivating application in consumer sales forecasting is introduced in Section~\ref{sec:app_dcmm}.  The case study context is supermarket sales of many individual items, and several examples of item-level sales highlight the uses of DCMMs and the flexibility of the new model class to adapt to substantially differing features of count time series.    As part of this, we discuss and develop a range of metrics for forecast assessment, including standard point-forecast measures, probabilistic calibration and coverage. The consumer forecasting field has tended to focus on very specific point-forecast metrics, and part of our work here is to broaden the perspective on forecast evaluation in response to, and enabled by, the availability of fully specific probability forecast distributions. 

Section~\ref{sec:multi-scale} addresses the interest in multivariate cross-series linkages and borrowing of information on shared characteristics and patterns. 
Our main focus here is on the potential for multivariate models to improve multi-step ahead forecasts at the level of individual series, while maintaining efficiency of the forward/sequential analysis and, critically, enabling scaling to many series.  Among multivariate approaches, a number of authors have explored models for counts or proportions~(e.g.~\citealp{Quintana1988};~\citealp{west1997book} chapter 16;~\citealp{DaSilva2016}). However,  there do not exist general classes of models addressing our key desiderata of flexibility at the single-series level, analytic tractability and capacity to scale to higher dimensions. Most existing models tend to be specific to applications and not easily amenable to integrating covariate information at the individual series level. Further, many require intense computations such as Markov chain Monte Carlo or sequential particle methods~\citep[e.g.][]{Cargnoni1997,Soyer2018}, which is antithetical to our concern for fast, sequential analysis and cross-series integration with many series.    Our work here builds on the flexible class of univariate DCMMs and defines a novel multi-scale approach to integrating cross-series information about common patterns, exemplified in terms of time-varying seasonality where a seasonal pattern is evident across series but with series-specific random effects. Critically, our new decouple/recouple approach enables information sharing while avoiding intense computations typical of random-effects/hierarchical models. The basic idea is of using aggregate level data to inform on micro-level series is one example of a decouple/recouple strategy that 
maximally exploits series-specific customization while enabling integration in multivariate models; see~\cite{ChenETALdynets2016JASA,GruberWest2016BA,gruber2017bayesian} for models that exploit this strategy in very different contexts. 
Section~\ref{sec:multi-scale_application} revisits application in the consumer sales case study, illustrating the use and impact of the 
multi-scale framework across several products in the context of sales forecasting of multiple related items. 

Additional comments in Section~\ref{sec:conc} and supporting technical material in the Appendix conclude the paper.

\section{Dynamic Count Mixture Models}\label{sec:model}

\subsection{Univariate Time Series and DGLMs \label{sec:basicDGLM} }

General notation follows that of standard Bayesian dynamic linear models~\citep{west1997book}. Denote by $y_t$ a univariate time series observed at discrete, equally-space times $t = 1, \ldots, T$.  At any time $t$ having observed $y_{1:t},$ available information is denoted-- and sequentially updated -- by 
$\cD_t = \{ y_t, \cD_{t-1},\cI_{t-1} \}$ where $\cI_{t-1}$ represents any relevant additional information becoming available at time $t-1$ in addition to past data (such as information used to define interventions in the model).  For any  vector of time indices $\seq{t+1}{t+k}$ for $k>0,$ forecasting $y_{\seq{t+1}{t+k}}$ at time $t$ is based on the information set $\{ \cD_t,\cI_t \}. $ 
Dynamic count mixture models (DCMMs) combine two examples of the broader class of 
dynamic generalized linear models (\citealp{west1985dynamic};~\citealp{west1997book}~chapter~15;~\citealp{Prado2010}~section~4.4).  The full class of DGLMs defines  Bayesian state-space modeling of data conditionally arising from distributions with exponential family form.
There are many benefits to Bayesian state-space modeling in this context. Data are modeled on their natural scale, and components of state-space models such as levels, trends, seasonality, and regression components are easily interpretable by non-statisticians. Time-varying model components allow non-stationarities to be captured and for models to adapt to unpredictable changes.  Analysis naturally implements sequential learning and forecasting through state evolutions and prior-posterior updates at each time $t$. The Bayesian framework allows incorporation of expert information or interventions into the model at any time point via modifications of \lq\lq current'' priors over state parameters.  Finally, Bayesian forecasting utilizes full predictive distributions rather than just point forecasts, and simulation at any time point provide facile access to summary forecasts on arbitrary functions of the future data over multiple steps ahead. 

In a specified DGLM,  the observation model has p.d.f. of exponential family form 
$p(y_t \mid \eta_t, \phi) = b(y_t, \phi)\exp{ [\phi \{y_t \eta_t - a(\eta_t)\} ]} $
where $\eta_t$ is the natural parameter that maps to linear predictor $\lambda_t = g(\eta_t)$ via link function $g(\cdot).$  Also $\phi$ is a known scale factor ($\phi=1$ in Bernoulli and Poisson models) and $b(\cdot,\cdot)$ a function specific to the chosen sampling distribution. 
Often, $g(\cdot)$ is the identity and the natural parameter is the linear predictor.  Dynamic  regression is defined by the state-space form
\begin{equation} \label{eq:DGLMregnevo}
\lambda_t = \F'_t \btheta_t\quad\textrm{where}\quad \btheta_t = \G_t \btheta_{t-1} + \bomega_t\quad\textrm{and}\quad \bomega_t \sim (\bzero,\W_t)
\end{equation} 
with the following elements: 
\begin{itemize} \itemsep=-3pt
\item  $\btheta_t$ is the latent, time-varying state vector and $\F_t$ is a known vector of constants or realized values of predictor variables (a.k.a. regressors).
\item The evolution equation specify a conditionally linear Markov process for the state vector through time: 
$\G_t$ is a known state matrix specifying structural evolution of the state vector, and  $\bomega_t$ is a stochastic innovation vector (or evolution \lq\lq noise''). 
\item The notation  $\bomega_t \sim (\bzero,\W_t)$ indicates that
$\text{E}[\bomega_t|  \cD_{t-1},\cI_{t-1} ]=\bzero$ and $\text{V}[\bomega_t | \cD_{t-1},\cI_{t-1} ] = \W_t$, the latter variance matrix being known at time $t-1.$  \item The $\bomega_t$ are independent over time and, at time $t-1$, $\bomega_t$ and $\btheta_{t-1}$ are conditionally independent given 
$\{ \cD_{t-1},\cI_{t-1} \}.$   
\end{itemize} 
Appendix~\ref{app:dglm}   details  forward filtering and forecasting analysis of DGLMs,  drawing 
on~\citealp{west1997book}~(chapter 15). This evolves information about state vectors over time based on partial posterior summaries in terms of 
mean vectors and variance matrices, with prior-posterior updates at each time based on linear Bayes' theory~\citep{GoldsteinWooff2007}.  
Further,  analysis exploits a variational Bayes method to constrain to conjugate priors for natural parameters at each time, so ensuring appropriate forms of implied forecast distributions. The combined linear Bayes/variational Bayes strategy avoids the need for additional assumptions and, more importantly, enables model analysis without resort to MCMC or other intensive computational approximations.  

DCMMs are based on two key special cases, those of  Bernoulli and Poisson DLGMs.   For a binary time series, changing the data notation to $z_t$ we have then $z_t \sim Ber(\pi_t)$,  Bernoulli with success probability $\pi_t.$ The traditional logistic DGLM has  $\lambda_t  = \text{logit}(\pi_t)$.     A conditionally Poisson DGLM for a non-negative count time series  $y_t$ has $y_t \sim Po(\mu_t)$ and natural parameter  
$\lambda_t = \log(\mu_t)$.   Conjugate analysis involves gamma priors for $\mu_\ast$, implying forecasts in terms of negative binomial distributions. 
More details of the analysis in these two cases appear in Appendix~\ref{app:dglm}, along with summary comments on the standard normal dynamic linear 
model~\cite[][chapter 4]{west1997book} that is also of interest in some applications.

\subsection{Flexible Mixtures of DGLMs: Dynamic Count Mixture Models \label{sec:dcmmmodel} }

A DCMM combines binary and conditionally Poisson DGLMs in a format similar to various existing models for time series of non-negative counts. It is often practically imperative to treat zero-versus non-zero outcomes separately from forecasting the integer outcomes conditional on them being non-zero. The novelty here is to use the flexible classes of DGLMs for the two components in an overall model, with dynamic predictor components in each that can be customized to context.  With non-negative count time series $y_t,$  define the binary series 
$z_t = \ind{y_t>0}$ where $\ind{\cdot}$ is the indicator function. A DCMM for outcomes $y_t$  is defined by observation distributions 
in which 
$$z_t \sim Ber(\pi_t) \quad\textrm{and}\quad 
   y_t | z_t = \begin{cases} 0, & \quad \text{ if }  z_t = 0, \\ 1 + x_t, \quad x_t \sim Po(\mu_t), &  \quad \text{ if } z_t = 1,\end{cases} 
$$ 
over all time $t.$ 
The parameters $\pi_t$ and $\mu_t$ are time-varying according to binary and Poisson DGLMs, respectively, i.e.,
\begin{equation} 
\text{logit}(\pi_t) = \F^0_{t} \bxi_t \quad\textrm{and}\quad  
\log(\mu_t) = \F^{+}_{t} \btheta_t \label{eq:dccmstates}
\end{equation}
with latent state vectors  $\bxi_t$ and $\btheta_t$  and known dynamic regression vectors $\F_t^+$  and $\F_t^0$, in an obvious notation. 
The regression vectors can include different model components if we expect different factors to impact the probability of a zero count and the size of the non-zero count. The conditional model for $y_t|z_t=1$ is  a shifted Poisson DGLM. In sequential learning, the positive count model component will be updated only when $z_t = 1$. That is, when a zero count is observed, the positive count value is implicitly treated as missing. This allows for a range of applications with a substantial probability of zeros over time. If, on the other hand,  a time series has few or no  zeros, the binary model will play a relatively limited role in forecasting. 

The forward filtering and forecast analysis evolves and updates prior moments of the state vectors in each of the binary and shifted Poisson model at each step, separately. Then, each forecasts one or more steps ahead by evolving state vector moments into the future, then applying the variational Bayes constraint to 
conditional conjugacy.  Thus, the marginal predictive distribution for $y_{t+k}$ at any $k>0$ steps ahead from time $t$ is the implied mixture of a Bernoulli and shifted Poisson, with the conjugate gamma prior predictive for the Poisson rate $\mu_{t+k}$ defining a conditional shifted negative binomial forecast distribution for that component.  In most applications, however, we are interested in full joint forecasts of paths $y_{\seq{t+1}{t+k}}$ over a sequence of future times $\seq1k$ from the current time 
$t.$  Looking at these joint predictive distributions provides information on dependencies between time points, and allows for calculation of other forecast quantities. For example, in forecasting daily sales of a supermarket item over each of the next $k=14$ days we may also be interested in quantities such as 
such the cumulative sales up to each day in that period,  or the number of those 14 days with zero sales, or the probability that cumulative sales exceed some specified level, and so forth.  To adequately (or at all) enable addressing such broader practical forecasting questions,  we forward-simulate the predictive distributions. That is, generate large Monte Carlo samples of the full predictive distribution $p(y_{\seq{t+1}{t+k}}|\cD_t,\cI_t)$ from the current time $t.$   
This is easily implemented as noted and detailed in Appendix~\ref{app:forecasting}, and such Monte Carlo samples can be trivially manipulated and interrogated to quantify forecast distributions for any function of the series of future outcomes of interest.

\subsection{DCMM Random Effects Extension}\label{sec:randomeffect}

DCMMs can flexibly model time series of counts with many or few zero counts. Another common characteristic of non-negative count data-- especially at higher levels of counts-- is over-dispersion relative to conditional Poisson models. The primary contextual development of the shifted Poisson DGLM will aim to 
customize the choice of $\F^{+}_{t}$ and associated evolution equation to best predict non-zero sales. While resulting forecast distributions may be generally accurate in terms of location, they may still turn out to under-estimate uncertainties and, in particular, fail to adequately capture infrequent extremes (typically higher, though sometimes also lower values of $y_t$).   Various approaches to this appear in the literature,  but all essentially come down to adding a representation of this excess and purely unpredictable variation. This is best addressed directly via random effects, and this is easily done in the DCMM using a novel random effects extension in the Poisson DGLM component. 

Start with  the shifted Poisson DGLM   with regression vector, $\F_{t,0}$, state vector, $\btheta_{t,0}$, and linear predictor $\F'_{t,0} \btheta_{t,0}$.  Call this the baseline model, i.e., the DGLM with no random effects. The random effects extension generalizes to the linear predictor $ \F'_{t,0} \btheta_{t,0} + \zeta_t$ where the $\zeta_t$ are time $t$-specific, independent, zero-mean random effects.  This is trivially implemented as an extended DGLM.  That is, redefine the state vector as $\btheta_t =  (\zeta_{t}, \btheta_{t,0}')'$ and the corresponding dynamic regression vector as $\F_t = (1, \F_{t,0}')'$, so that the new model has 
$\log(\mu_t) = \F'_t \btheta_t = \F'_{t,0} \btheta_{t,0} + \zeta_t$ as required; this simply defines a different, and more general DGLM that admits time $t$ individual and unpredictable variation over and above the baseline.     The state evolution equation will be modified to add a first row and column to the state evolution matrix with zero elements representing lack of dependence of random effects time-to-time as well as independence of other state vector elements.  It remains to specify levels of expected contributions of random effects, and this is done using a random effects discount factor $\rho$, building on the standard use of discount factors for DGLM evolution variance matrices as a routine~(Appendix~\ref{app:discountfactors} here; see also 
\citealp{west1997book} chapter 6).  In particular, at each time 
$t-1$ suppose that prior uncertainty about the core state vector elements $\btheta_{t,0}$ at time $t$ is reflected in the prior variance matrix $\R_{t,0},$ so that 
the uncertainty about the baseline linear predictor is represented by 
$q_{t,0} \equiv \text{V}[\F'_{t,0}\btheta_{t,0}| \cD_{t-1},\cI_{t-1}] = \F_{t,0}'\R_{t,0}\F_{t,0}.$ Then, a random effects discount factor $\rho\in (0,1]$ 
defines the conditional variance of $\zeta_t$ by 
$v_t \equiv \text{V}[\zeta_t| \cD_{t-1},\cI_{t-1}] = q_{t,0}  (1-\rho)/\rho.$ 
If $\rho$ is set to one, then this model is simply the Poisson DGLM without the random effect. As $\rho$ gets closer to zero, the variance of the random effect increases. Here $\rho$ becomes a model hyper-parameter to be explored along with others.   For $\rho<1,$ the additional variance injected into the time $t$ prior is now relative to the variance of the underlying baseline models, so we have access to interpretation of $\rho$ as defining a relative  or percentage contribution to predictive uncertainty, as with standard discounting in state-space models.   Then, the impact is seen in increased dispersion of forecast distributions; some aspects of this on increased variance of the predictive negative binomials for future non-zero counts are highlighted in further technical details in 
Appendix~\ref{app:discountrandomeffects}.


\section{Product Sales Forecasting with DCMMs}\label{sec:app_dcmm}

\subsection{Context and Data}

\begin{figure}[ht!]
\centering
\begin{tabular}{c}
\includegraphics[width = .65\textwidth]{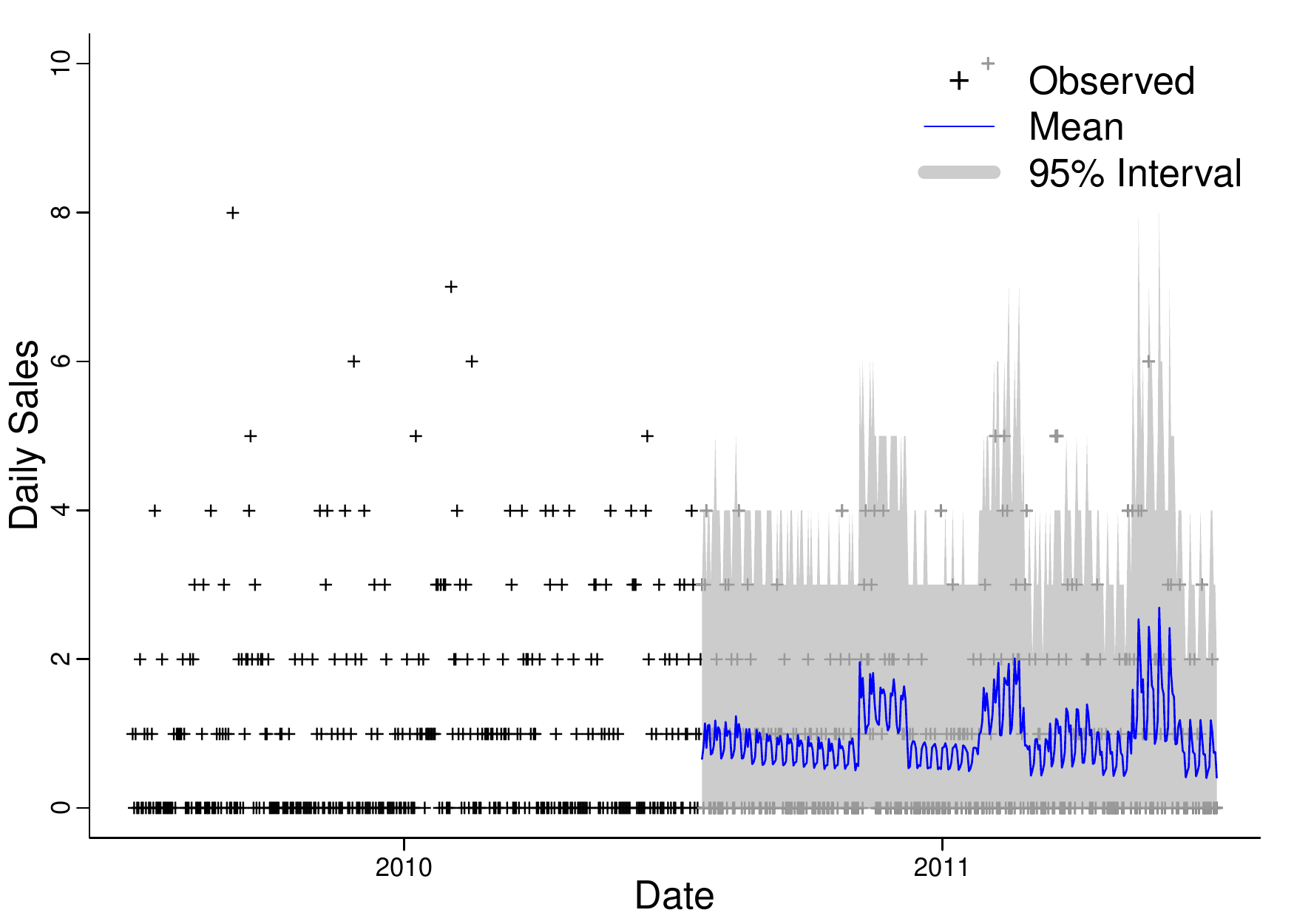}\\
\includegraphics[width = .65\textwidth]{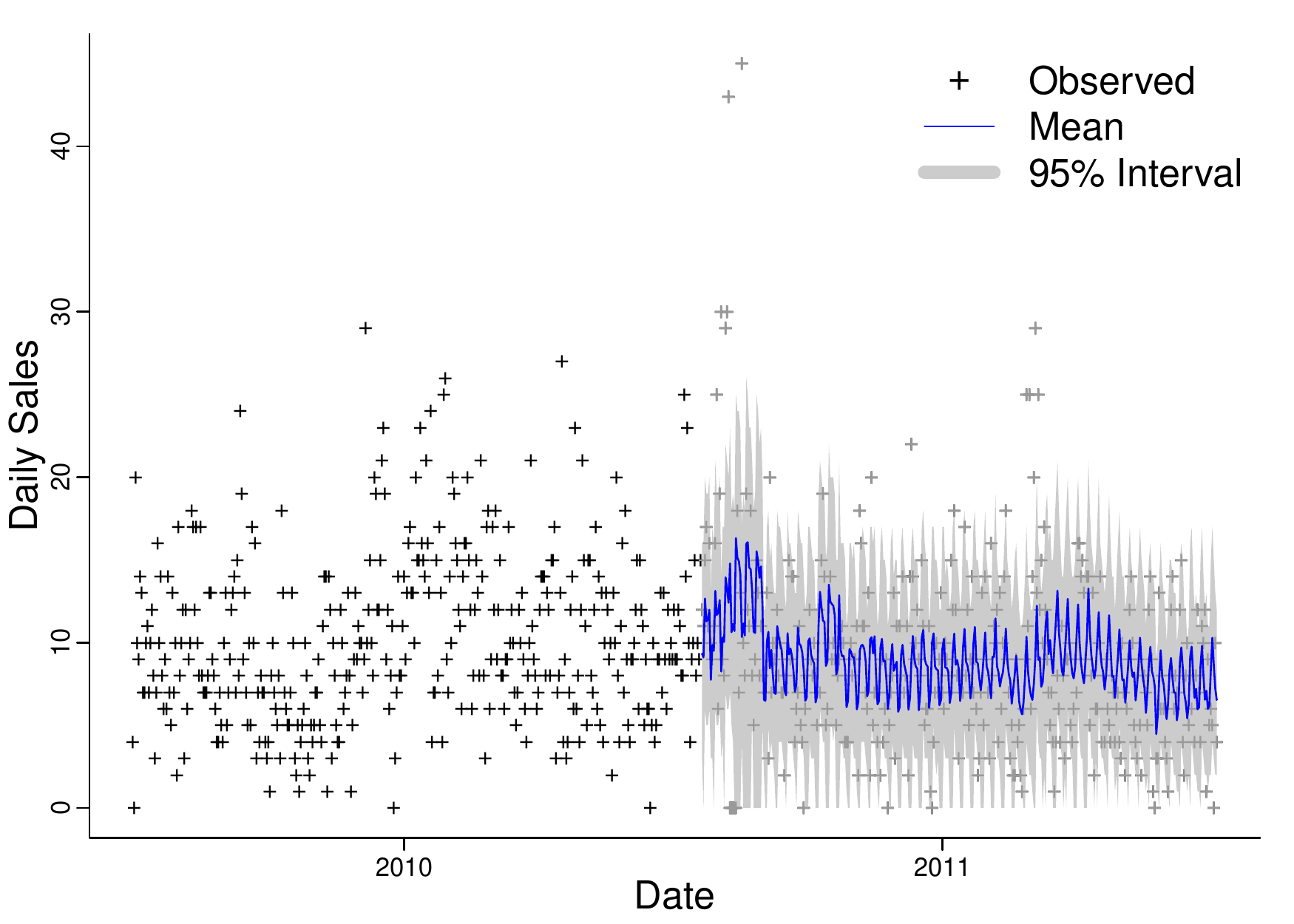} 
\end{tabular}
\caption{Data and aspects of $1-$step ahead forecast distributions for items A (upper) and B (lower).  Shading:  95\% predictive credible intervals; full line:  predictive mean.}\label{fig:dcmmforecasts}
\end{figure}

Our case study concerns multi-step forecasting of many individual items in each of a large number of US supermarkets (or stores). An item is defined by a unique stock keeping unit (SKU) and sales forecasts are required to be updated daily for the following $\seq1{14}$ days.  For our examples here we extract a very small number of items at one chosen store to provide illustrations and insights into the utility of DCMMs. 
The selected data set records daily sales of 179 SKUs in one particular store over the 2{,}192 days from
July~1st~2009 to July~1st~2015. Each of these products is in the pasta category, in one of 14 subtypes of pasta. By percentage of unit sales, the primary pasta types are spaghetti (25\%), macaroni (13\%), wholewheat (10\%), and penne (9\%). The products comprise 20 brands, and the majority of unit sales (44\%) come from the supermarket's in-house brand. Additional information includes the price paid per transaction, and whether or not each SKU was on promotion on the date of the transaction. In the contexts of analyses to follow, the data for several items are summarized in Table~\ref{tab:pastasummaries} and shown in Figures~\ref{fig:dcmmforecasts}~and~\ref{fig:ms_items}; these provide some insights into heterogeneity of daily sales patterns. 
\begin{table}[htbp!]
\centering
\begin{tabular}{lccccc}
\hline
Item & Mean  &Median & Variance & \% 0 sales \\
\hline
A &  0.9 & 0 & \ph 1.8 & 52.3 \\
B &  9.6  & 9  & 30.2 & \ph 1.6\\
C & 9.5  & 9 & 27.4 & \ph 1.2 \\
D &  3.2  & 2 & \ph 9.6 & 15.2 \\
E &  2.4  & 2 & \ph 5.5 & 19.2\\
F & 0.1 & 0 & \ph 0.2 & 90.7\\
\hline
\end{tabular}
\caption{Some summaries of daily pasta sales data by item} 
\label{tab:pastasummaries}
\end{table}

\subsection{Example Univariate DCMMs}
As example series, data in Figure~\ref{fig:dcmmforecasts} show daily sales of two selected spaghetti items in one store over the two year period from July~1st~2010 to July~6th~2011. 
Item A has relatively low daily sales with a mean of 0.9, a median of 0, and zero sales occurring on 52\% of days. Based on the prevalence of days with zero sales, the binary component of the DCMM will play an important role in forecasting item A. Item B is the highest selling spaghetti product in this store, with a median of 9 sales per day and zero sales occurring on only 1.6\% of days. The daily sales of item B appear to have high variance, and on some days we see sales spike to above 40 units. 

The same form of DCMM is applied to each item.  Each of the Bernoulli and shifted Poisson components has a local level, a regression component with scaled log price as predictor, and a full Fourier seasonal component of period 7 to reflect the day-of-week effects. All components are dynamic, allowing for variation over time in local level, regression coefficient, and Fourier coefficients.  Thus each of the binary and Poisson DGLMs have regression vector and evolution matrix defined by
$$\F'_t = \begin{pmatrix} 1, \, \log(\textrm{price}_t), \, 1,0, \, 1,0, \, 1,0 \end{pmatrix}
\qquad\textrm{and}\qquad
   \G_t = \text{blockdiag}[1,\, 1,\, \H_1,\, \H_2,\, \H_3]
$$ 
with
$$\H_{j} = \begin{pmatrix} \phantom{-}\cos(2\pi j/7)  & \sin (2\pi j/7) \\
-\sin (2\pi j/7) & \cos(2 \pi j/7)\end{pmatrix},\quad j=\seq13.$$
Exploratory analysis of an initial three weeks of data was used to specify prior moments on the state vector at $t=1$ representing day 22 of the full data set.  For the Poisson component,  this used standard reference Bayesian analysis assuming no time variation in parameters to define \lq\lq ballpark'' initial priors.   For the Bernoulli component, the level has prior mean $\log(p/(1-p))$ where $p$ is the proportion of the first 21 days on which a sale occurred, with all other prior means set to zero and the prior variance matrix set as the identity. Fixed discount factors of 0.999 (Bernoulli) and 0.99 (Poisson) are used on each of the level, regression and seasonal components; these values were  chosen based on the results of previous analyses of daily sales data. 
The DCMM analyses were run through the first year of data before forecasting. From the start of the second year, full forecast distributions for 
$\seq1{14}$ days ahead were computed at each day, updated recursively to the end of the year.  These define Monte Carlo samples of size 5{,}000 of synthetic future sales over rolling 14 day periods.  Some illustration of these forecasts appears in Figure~\ref{fig:dcmmforecasts}.

\subsection{Point Forecasts and Metrics}\label{sec:eval_counts}

The consumer sales/demand forecasting and other literatures cover a range of metrics for forecast evaluation, with variants of traditional loss functions for point forecasts customized to count data with concern, particularly, for cases of low counts~\citep[e.g., among recent contributions, see][]{kolassa2016evaluating,hyndman2006another, fildes2007against,yelland2009bayesian,gneiting2011making,morlidge2015measuring}. As noted in~\cite{hyndman2006another}, simply adopting common measures of forecast accuracy can produce \lq\lq misleading results" when applied to low valued count data. 

Our view is that \lq\lq a forecast'' is the full predictive distribution rather than one or more point summaries. For actionable decisions, understanding the potential implications of uncertainty as reflected in the full distribution can be key, while also adding significantly to evaluation and comparisons of forecast accuracy. Any point forecast selected should be rationalized and understood as a decision made on the basis of utility/loss considerations in the forecasting context, with implicit or explicit derivation from a decision analysis perspective. The predictive mean is optimal under squared error loss, the median for absolute error loss, and the mode for the (typically not substantively relevant) 0-1 loss. If the loss function is an asymmetric piecewise linear function, 
$L(f,y) = (\ind{f\geq y} - \alpha)(y - f)$ for outcome $y$ with point forecast $f$, then the $\alpha$-quantile of the predictive distribution is the optimal point forecast.  Modifications of the absolute percentage error (APE) loss function $L(f,y) = |1- f/y|$
are commonly used in consumer demand forecasting of strictly positive counts $y$. Under this loss, the optimal $f$ is  the $(-1)$-median, i.e., the median of the p.d.f. $g(y) \propto p(y)/y$ when $p(y)$ is the forecast p.d.f., although typical application is based on sub-optimal choices of $f$ as full forecast distributions are rarely developed.  It is also easy to explore novel modifications and extensions of loss functions from a decision analysis perspective. For example, APE
does not allow for zero outcomes, while practical extensions-- such as ZAPE, with  $L(f,y) = |1- f/y| \ind{y>0} + l(f) \ind{y=0}$ for some increasing function $l(f)>0$-- are amenable to simple optimization to define relevant point forecasts if desired.   Specific loss functions should be chosen in the context of resulting decisions to be made. In inventory control, there are costs associated with missed sales due to stock-outs, as well as the cost of overstocking items. The forecaster may be interested in a quantile of the distribution to reflect these utilities. 

To  connect with common practice and recent literature, we explore various point forecast metrics in Section~\ref{sec:multi-scale_application}. 
First, however,  we broaden perspective to evaluation of the full forecast distributions using the practically relevant issues of coverage and calibration of predictive distributions.

\subsection{Probabilistic Forecast Evaluation: Examples in Sales Forecasting}\label{sec:eval_probs}

\begin{figure}[hp!]
\centering
\includegraphics[width = .3\textwidth]{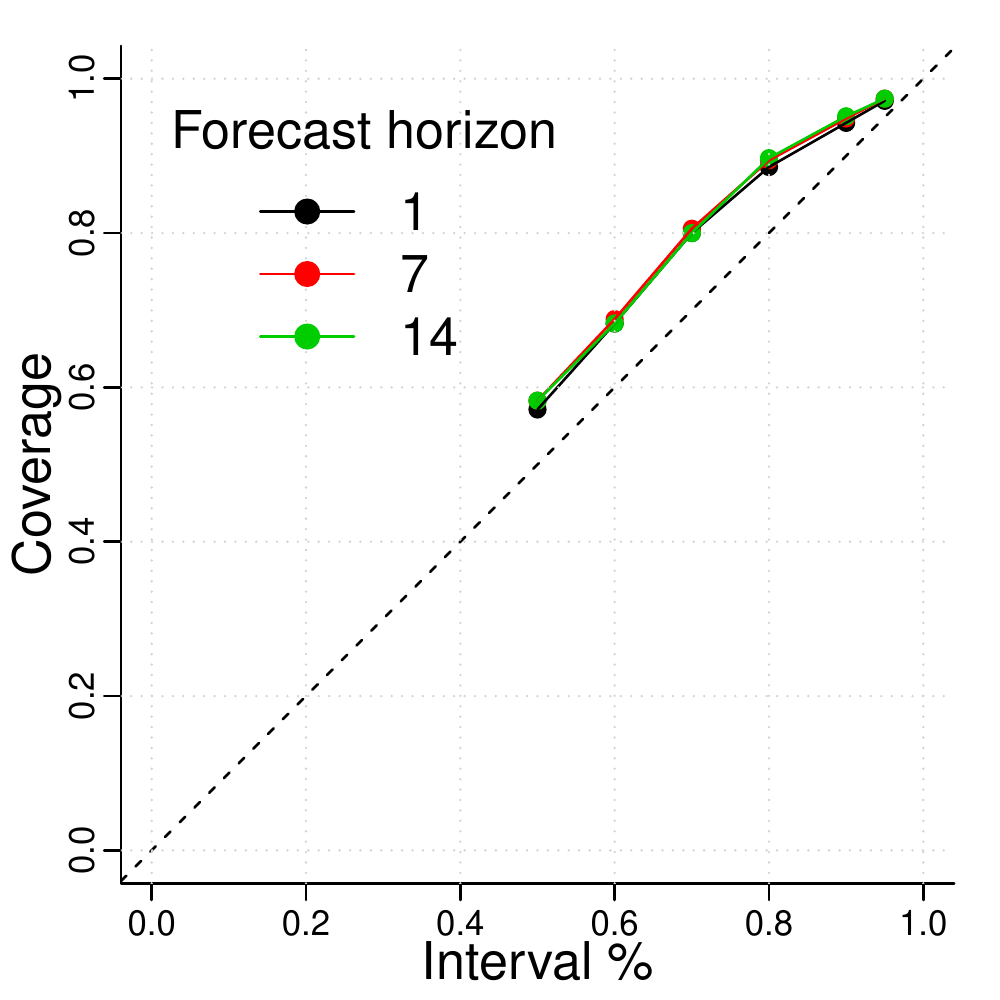} 
\qquad
\includegraphics[width = .3\textwidth]{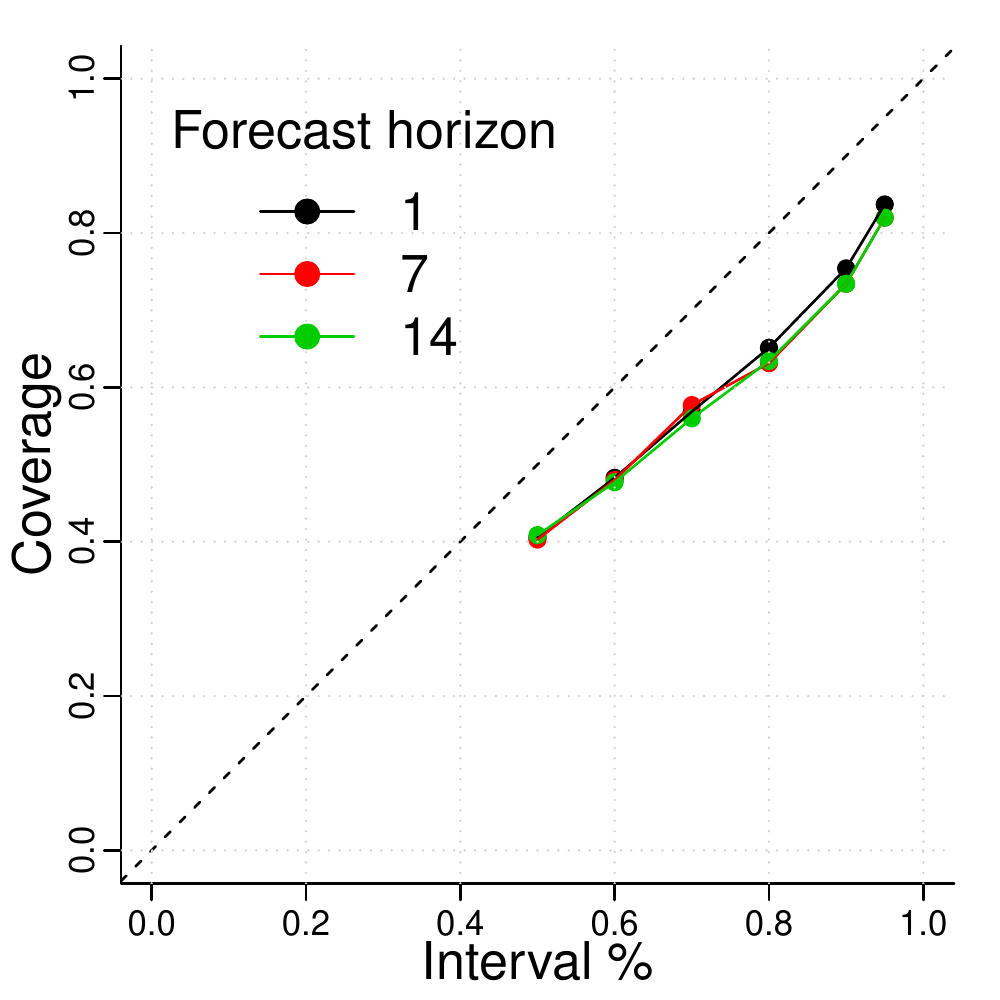} 
\caption{Coverage plots for items A (left) and B (right) from $1-$, $7-$ and $14-$day ahead forecasts. }\label{fig:dcmmcoverage}
\includegraphics[width = .3\textwidth]{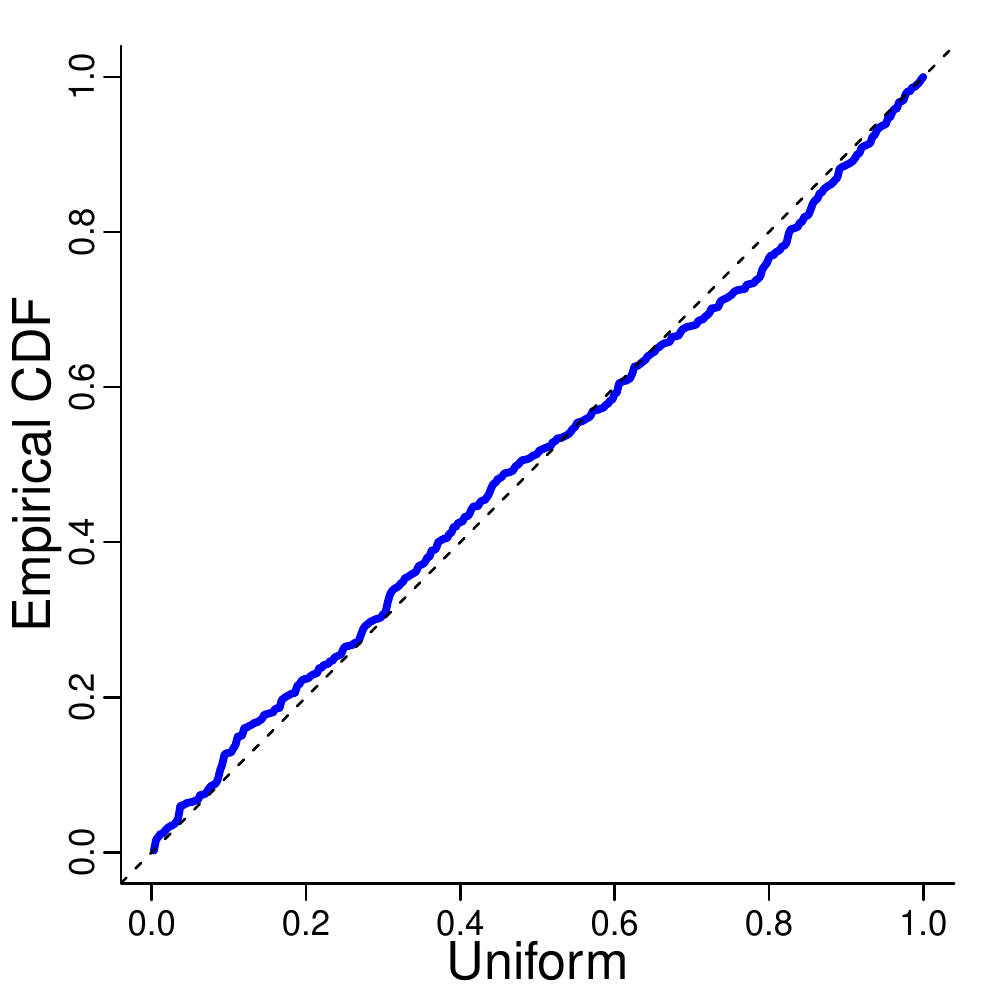} 
\qquad
\includegraphics[width = .3\textwidth]{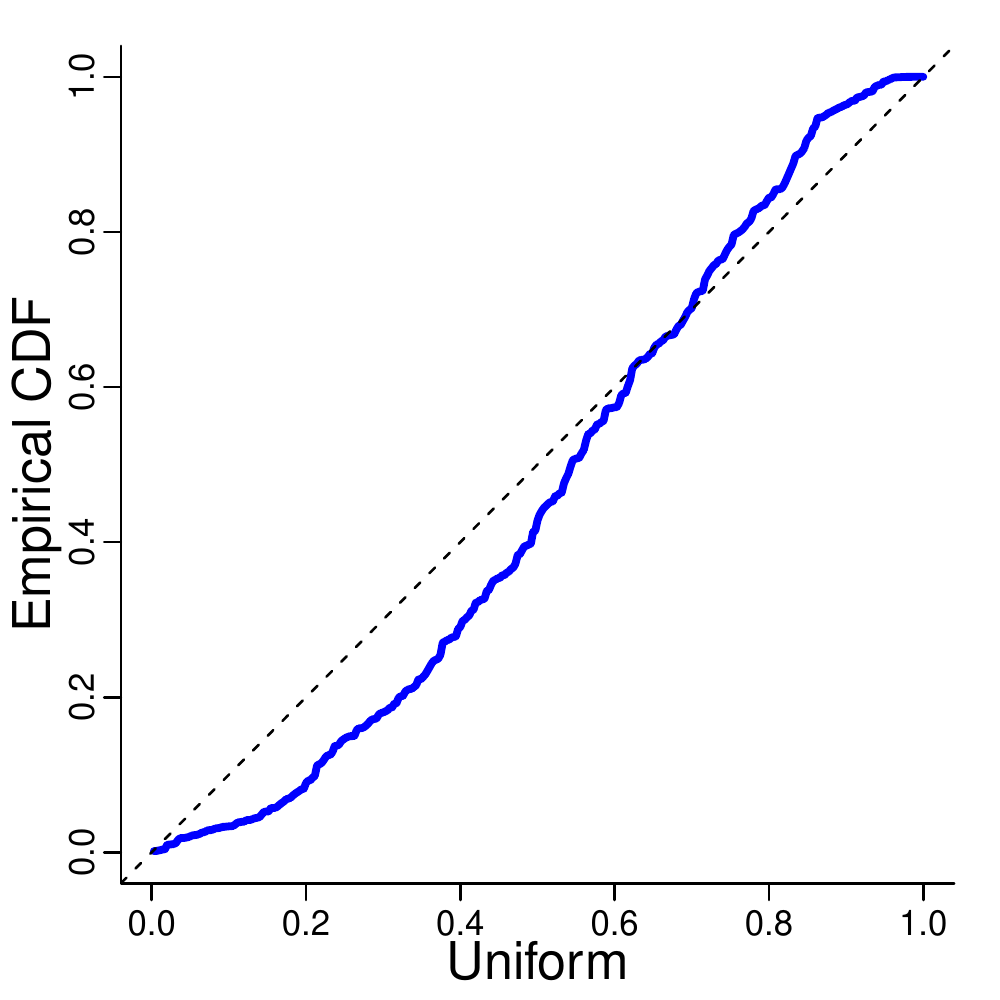} 
\caption{Randomized PIT plots from $1-$day ahead forecasting of items A (left) and B (right). Full line: ordered randomized PIT values; dashed:  $45-$degree line.}
\label{fig:dcmmcalibration}
\includegraphics[width = .3\textwidth]{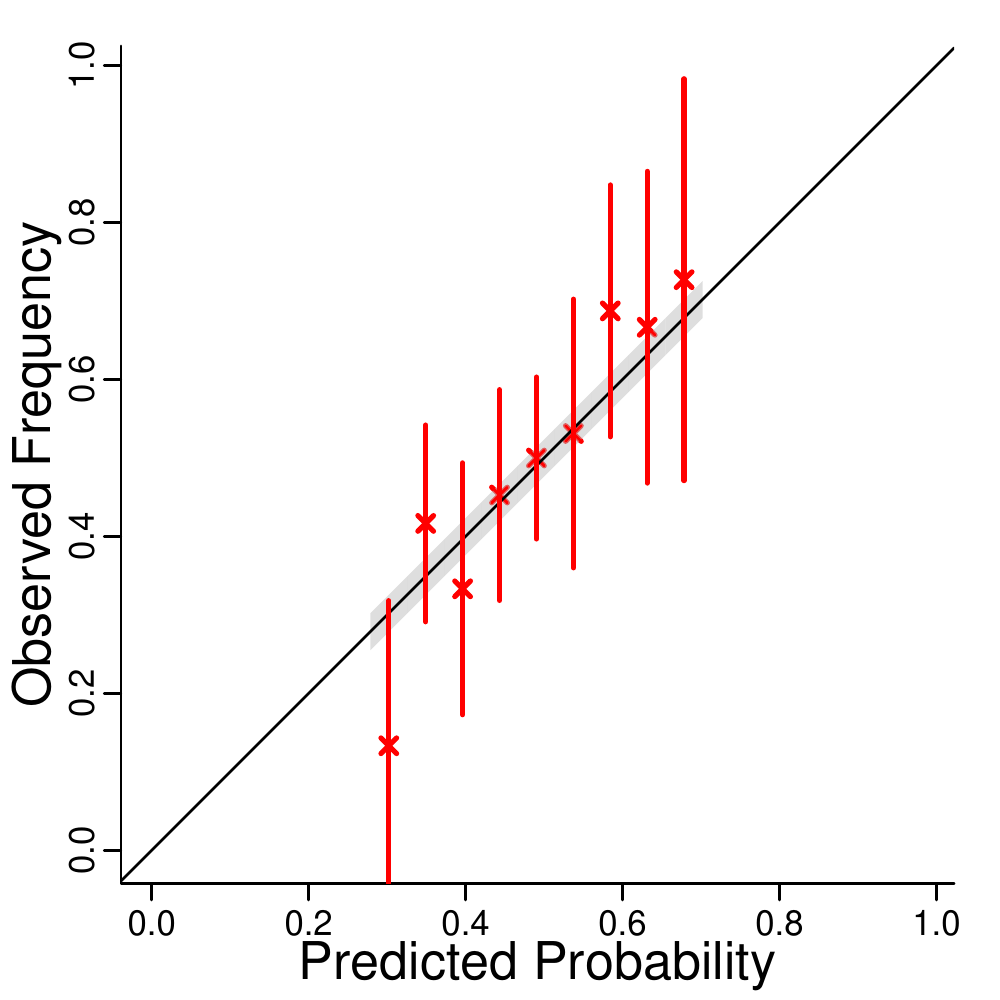} 
\qquad
\includegraphics[width = .3\textwidth]{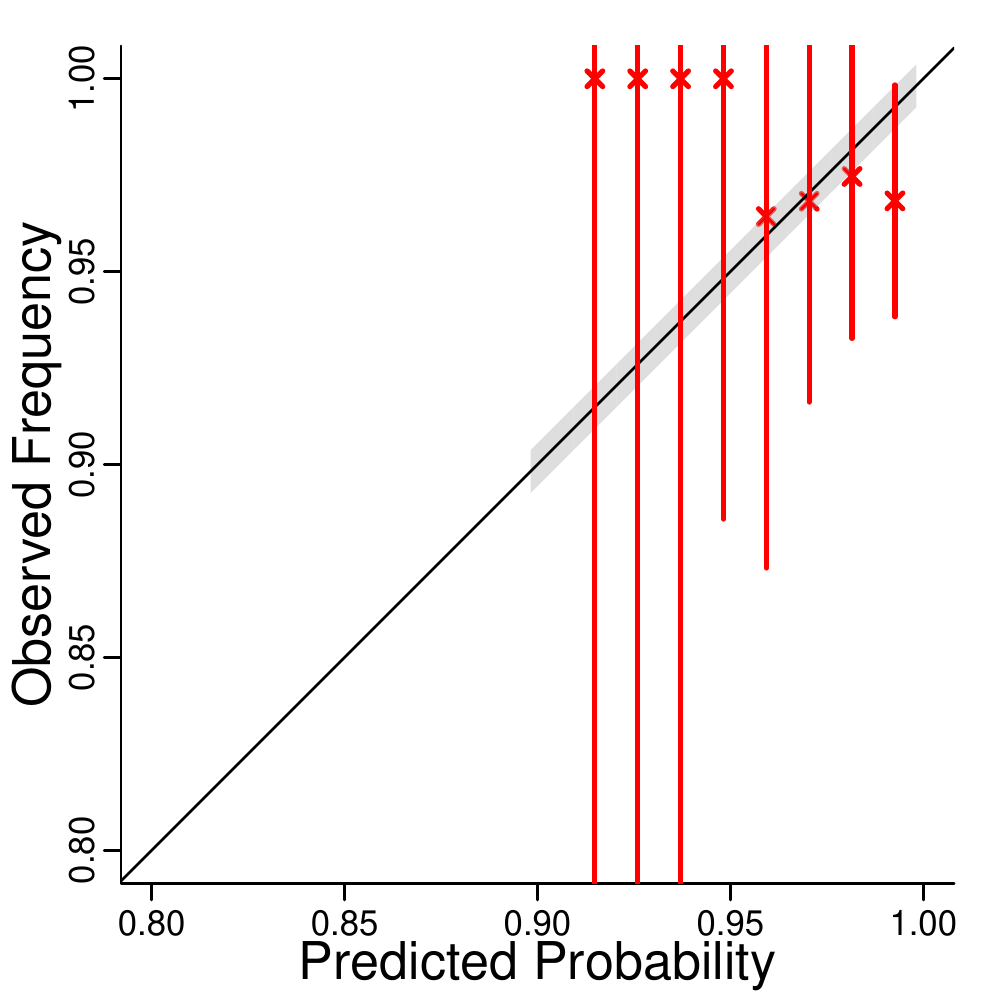} 
\caption{Binary calibration plots from $1-$day ahead forecasting of non-zero sales  of items A (left) and B (right). Crosses mark observed frequencies in each bin, horizontal grey shading indicates variation of forecasts in each bin, and vertical  bars indicate binomial variation based on the number of days in each bin.}
\label{fig:dcmmbinary}
\end{figure}

Figure~\ref{fig:dcmmforecasts} gives the overall impression that the DCMM forecasts relatively well in the short-term for both items A and B, clearly picking-up the seasonal patterns and responding to changes over time.   Then, the infrequent higher sales levels are better forecast for item A than apparently for item B, the latter being a higher-selling item. To explore forecast performance in more detail, 
Figures~\ref{fig:dcmmcoverage},~\ref{fig:dcmmcalibration} and~\ref{fig:dcmmbinary} summarize  aspects of the full predictive 
distributions generated by the DCMMs for item A (left frames) and B (right frames).   

Figure~\ref{fig:dcmmcoverage} displays coverage of forecast distributions for each of 1, 7, and $14-$days ahead.  These graph the 
empirical coverage obtained over the full year of forecasting for predictive credible (highest predictive density-- HPD) intervals in each case. 
An ideal model would lead to coverage plots close to the $45-$degree line.  For item A,  forecast distributions have slight over-coverage. For example, empirical coverage of the $1-$day ahead 50\% predictive intervals is about 57\%. In contrast, for item B we see evidence of 
under-coverage at all horizons, related mainly to the apparent inability of the model to adequately forecast the infrequent higher sales values. 
 
A second probabilistic evaluation uses the probabilistic integral transform (PIT), i.e., a general residual plot based on the predictive c.d.f. for each outcome.  
Since predictive distributions are discrete, this involves the randomized PIT~\citep{kolassa2016evaluating}.   If sales counts $y$ are forecast with predictive c.d.f. $P(\cdot)$, define $P(-1) = 0$ and draw a random quantity $p_y \sim U(P(y - 1), P(y))$ given the observed value of $y.$  
Over the sequence of repeat forecasting events an ideal model would generate realizations 
 $p_y$ that are approximately uniform. For each item, Figure~\ref{fig:dcmmcalibration} plots the ordered randomized PIT values from the $1-$day ahead forecasts distributions versus uniform quantiles.  The concordance of the outcomes with uniformity is apparently strong for item A.  For item B, however, we see significant non-uniformity and a shape consistent with forecast distributions that are just too light-tailed; that is, the outcome sales data on item B exhibit higher levels of variation than the DCMM predictive distributions capture. 
 
The third  probabilistic evaluation focuses on frequency calibration properties of forecasts of zero/non-zero sales.  Ideal calibration means that, of the days non-zero sales are forecast with probability near $p$,  approximately $100p\%$ actually have non-zero sales.  In practice, we bin the probability scale according to  variability in forecast probabilities of non-zero sales across the year, and evaluate the realized frequency of non-zero sales on the days within each bin. Figure~\ref{fig:dcmmbinary} displays the results for  $1-$step ahead binary predictions. 
For item A, the predicted probabilities of non-zero sales range in $\seq{30}{80\%};$ these are allocated into ten bins of equal width. The figure displays the observed frequency of non-zero sales within each bin and an approximate 95\% binomial confidence interval based on the number of days within each bin. Horizontal shading displays the width of the predicted probability bins. Ideal predictions would lead  the observed frequency in each bin to fall within the shaded area, while the vertical bars indicate limits based on sample size in each bin. For item A, the performance is apparently very good indeed. 
For item B, predicted probabilities of  non-zero daily sales range in $\seq{90}{100\%}$ over time; this is a relatively high-selling item. 
As with item A, the vertical calibration intervals intersect the $45-$degree line and the horizontal shading, indicating that there are no obvious issues with forecasting the zero/non-zero outcome. In terms of the full DCMM, the binary component  is obviously less important for item B than for item A. Then, as noted above, the under-dispersion of forecasts of item B is a clear negative; this is addressed with the random effects extension below. 
 
\subsection{Random Effects Extension}\label{sec:app_re}

\begin{figure}[htb!]
\centering
\includegraphics[width = .65\textwidth]{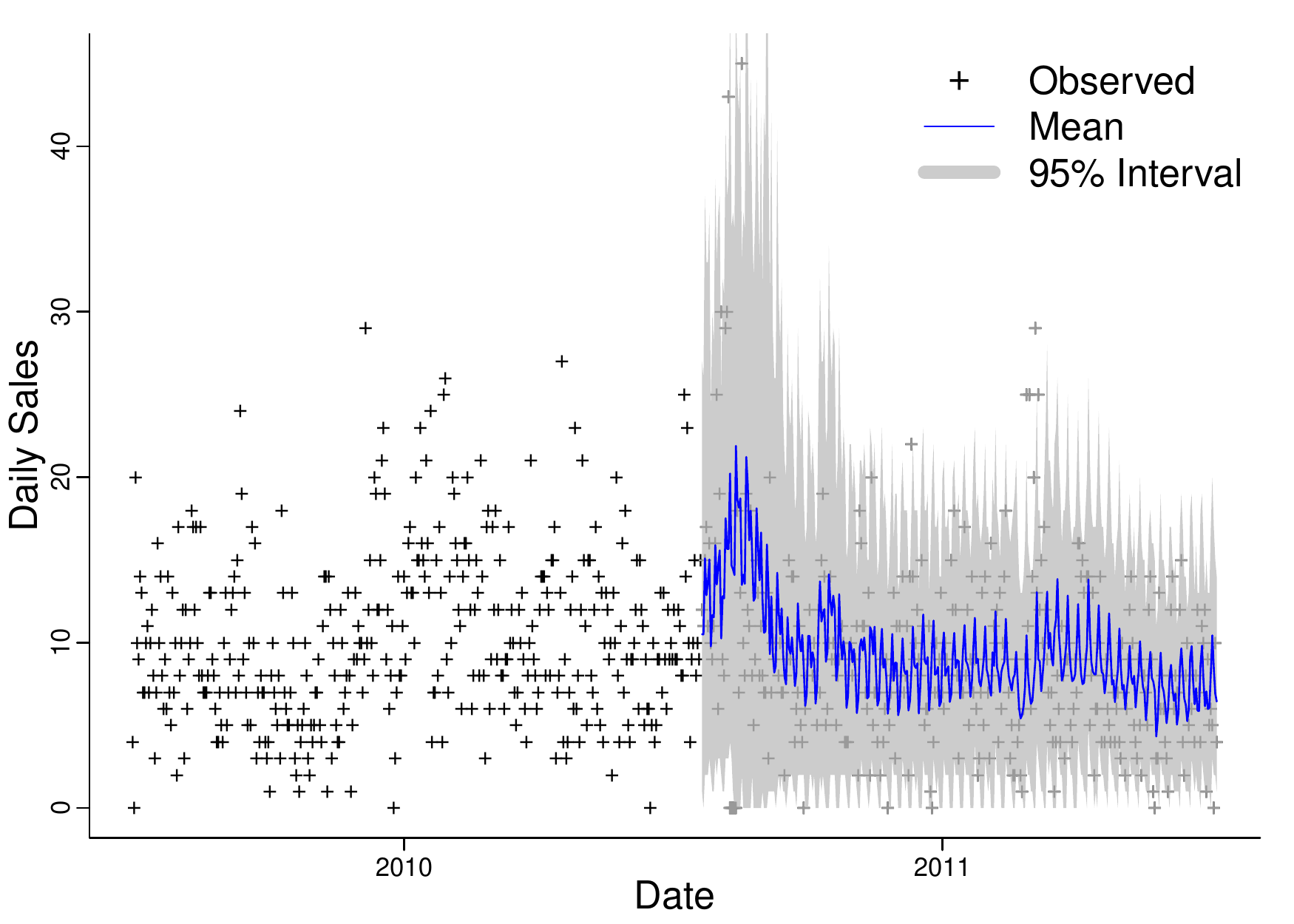}
\caption{Daily sales of item B, and the $1-$day ahead forecast from the DCMM with random effects. The blue line indicates the forecast mean, and the gray shading indicates the forecast distribution 95\% credible interval.}\label{fig:REforecast}
\end{figure}

\begin{figure}[ht!]
\centering
\includegraphics[width = .3\textwidth]{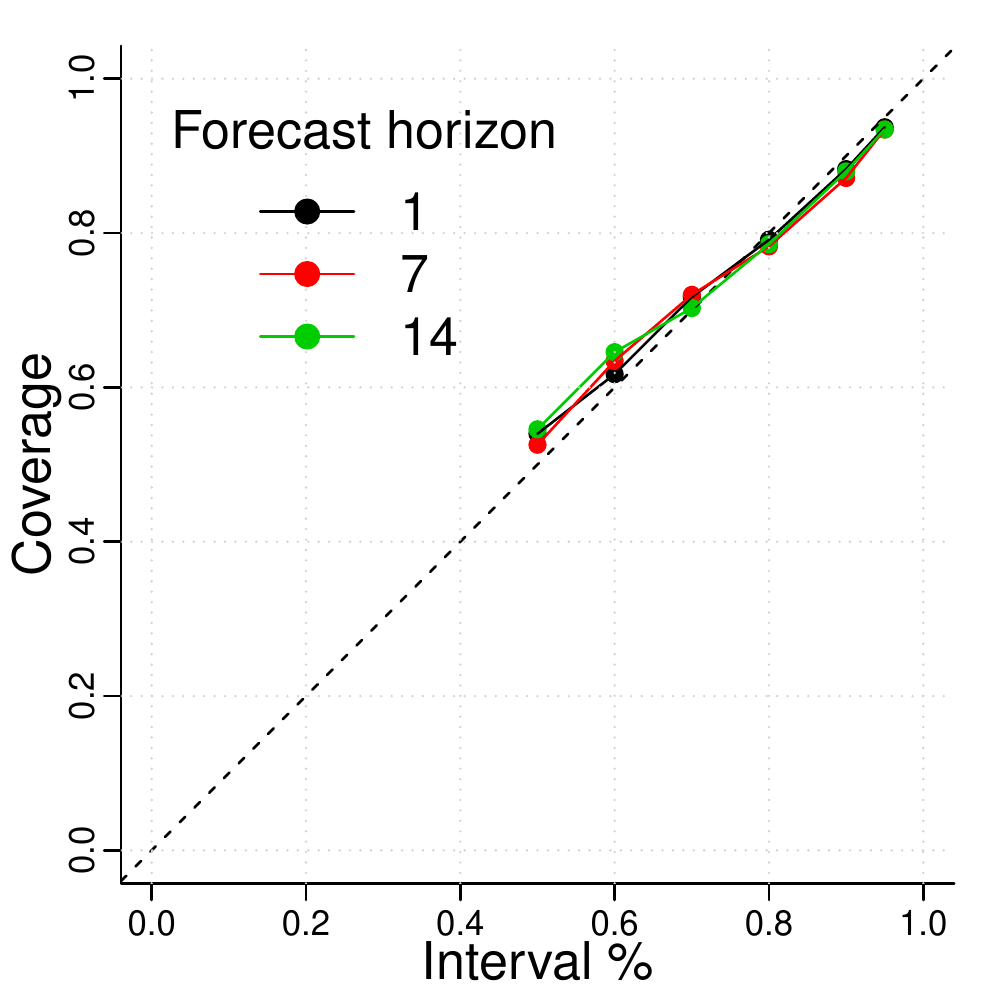} 
\qquad
\includegraphics[width = .3\textwidth]{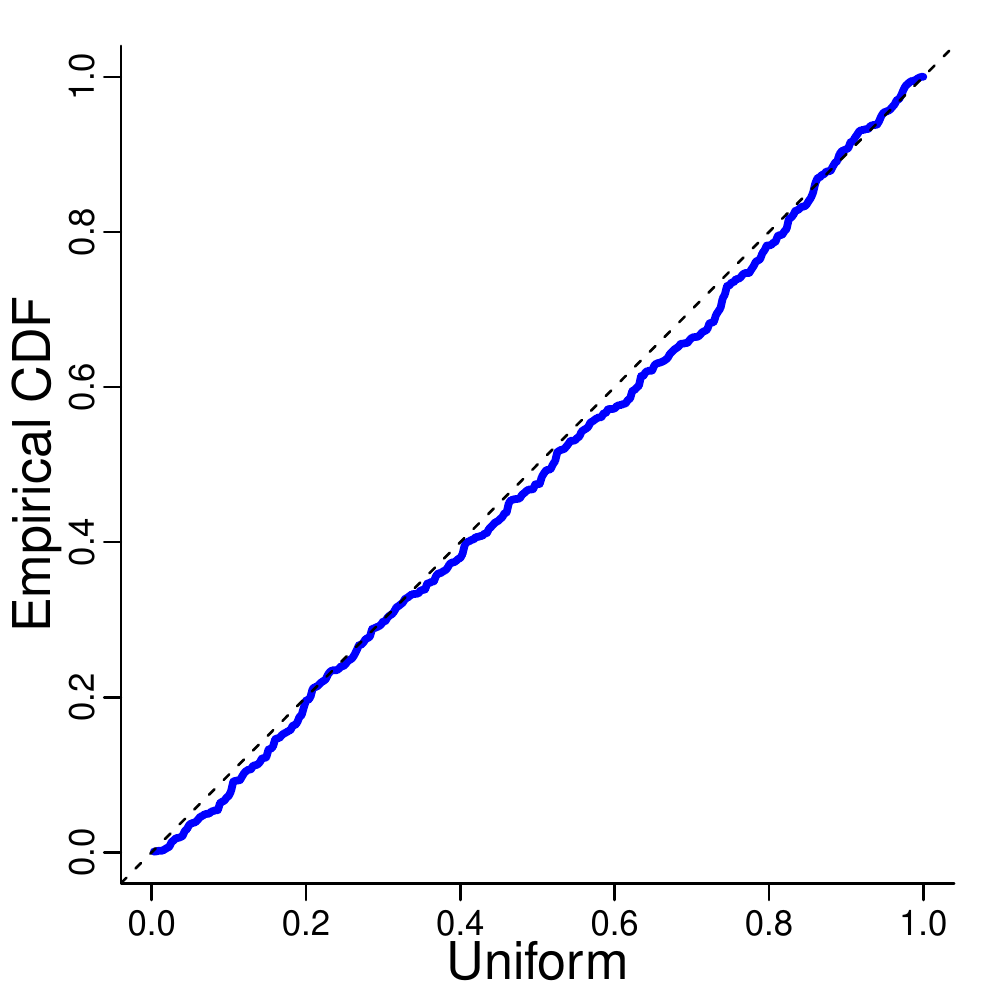} 
\caption{Empirical coverage plot (left) and randomized PIT plot (right) for $1-$day ahead forecasting of item B using the DCMM with random effects extension.  Compare with the results under the basic DCMM in the right-hand frames of Figures~\ref{fig:dcmmcoverage} and~\ref{fig:dcmmcalibration}.}
\label{fig:REresults}
\end{figure}

We illustrate the potential of the random effects extension of DCMMs introduced in Section~\ref{sec:app_dcmm} to adapt to the over-dispersion issues in the basic analysis of item B in the last section.  The main details of the analysis remain the same, but now the model is modified to include day-specific random effects. The summarized analysis uses a random effects discount factor $\rho = 0.2$, chosen following exploration of the impact of different values across the first year of data. Figure~\ref{fig:REforecast} displays the updated $1-$step forecast means and 95\% credible intervals over time, to be compared to Figure~\ref{fig:dcmmforecasts}.  Note the  wider forecast intervals that are to be expected. Figure~\ref{fig:REresults} displays the resulting coverage of the 1, 7, and $14-$day ahead forecast credible intervals over time, and the calibration plot of the randomized PIT values from $1-$day ahead forecasts. Coverage has increased and substantially improved to conform with the $45-$degree line, while PIT values are in much closer concordance with uniformity.  Overall, the addition of the random effect has accounted for some of the under-dispersion of forecast distributions in the baseline DCMM, and these aspects of forecast evaluation indicate clear and practical improvements as well as overall accuracy. 

\section{Multivariate Forecasting Framework}
\label{sec:multi-scale}

\subsection{Decouple/Recouple and Shared Features Across Series} 
We now turn to recoupling sets of univariate DCMMs in contexts where there may be opportunity to improve multi-step forecasts via more accurate assessment of effects and patterns shared across the series.   That is, we aim to exploit traditional ideas of hierarchical random effects models where common features over time are \lq\lq seen differently'' by each univariate series, and where both series-specific effects and noise obscure the patterns at the individual series level. This is particularly relevant in sales and demand forecasting when items are sporadic, i.e., in our DCMMs in cases of non-negligible zero sales and otherwise low count levels.  Products can often be grouped hierarchically based on characteristics like product family, brand, or store location.  Sales and demand patterns within such groups may have similar trends or seasonal patterns due to external factors such as marketing, economy, weather, and so forth. Our main example here focuses on the daily seasonal pattern over each week, a pattern that is heavily driven by customer traffic through the stores related to weekly behavioral factors.  This general pattern is naturally shared by many individual series, but at low levels of sales it is substantially obscured by inherent noise so that using information from other series distilled through a multivariate approach is of interest.  If products are grouped we may expect estimates of seasonality at the aggregate level to be more accurate and less noisy than at the individual level. Using aggregated seasonality in a model rather than individual item seasonality may then improve forecasting for individual items. 
On the other hand, seasonality exhibited by items that sell at higher levels may be much more evident and the potential to gain forecast accuracy there less obvious. Part of our interest is thus to explore the potential across a range of items.   

Our desiderata include maintaining flexibility in customizing DCMMs  to individual time series along with the ability to run fast, sequential Bayesian analysis of  inherently decoupled univariate analyses of many series.  In product demand forecasting, retailers are generally interested in many items (often thousands) simultaneously. Due to time and computational constraints, the conventional approach is to rely on univariate methods which forecast independently across SKUs and therefore this is a central consideration.  That is, we avoid large-scale and complex multivariate modeling that would otherwise necessitate the use of  intense MCMC (or other) computations that would obviate the use of efficient sequential analysis while most seriously limiting the ability to scale in the number of time series. To do this, we maximally exploit the univariate framework of Section~\ref{sec:model} with series  decoupled conditional on factors shared in common across series, and then recouple by utilizing a separate model to analyze and forecast these common factors. 

\subsection{Multi-Scale Models and Top-Down Recoupling}
 
The essential structure is that of a set of decoupled dynamic latent factor models as relates to traditional Bayesian multivariate approaches for conditionally normal data~\citep[e.g.][]{aguilar1999bayesian,aguilar2000bayesian,carvalho2007dynamic}, but with the major novelty of inferring latent factors externally.  The latter draws conceptually on prior work in multi-scale models in time series~\citep[e.g.][]{Ferreira2003,Ferreira2006} and on \lq\lq bottom-up/top-down'' ideas in Bayesian forecasting~\citep[][section 16.3]{west1997book}.   The basic idea of top-down forecasting is to forecasting patterns at a highly aggregate level
and then somehow  disaggregate down to the individual series. Practical methods using point forecasts of aggregate sales over a set of items typically use some kind of weighting to disaggregate. While our framework is general, we focus on the key example of seasonal patterns, where our work contributes to an existing literature. For example, the method of group seasonal indices (GSI:~\citealp{withycombe1989forecasting}; \citealp{chen2007use}) aggregates  series in a defined group and then estimates seasonal factors from the aggregate series. We follow this concept, but with an holistic class of Bayesian state-space models-- with time-varying patterns and full probabilistic specifications-- rather than direct data adjustments assuming constant seasonal patterns.  The full Bayesian approach was mooted in conditional linear, normal examples in~\citep[][section 16.3]{west1997book}, and the developments here extend that to a full class of multivariate DCMMs for count series.  Importantly, we use full simulation of posterior distributions of common factor patterns to send to the set of univariate DCMMs, so that all relevant  uncertainties are quantified and reflected in the resulting forecast distributions of the 
univariate series. 

\subsection{Model Structure and Analysis}

\subsubsection{Notation and Structure} 
A set of $N$ series  $y_{i,t}$,  $(i =\seq1N),$ follow individual DCMMs   sharing some common factors of interest.  Denote these individual models by $\cM_i$ for $i=\seq1N.$  In each of the binary and shifted Poisson DGLM components,  $\cM_i$ has both series-specific and common state-space components with latent factor predictors shared across the $N$ series.  
In $\cM_i$  the time $t$ state and regression vectors for each DGLM component have partitioned forms with individual and common predictors. For example, the shifted Poisson component has state and regression vectors defined by 
\begin{equation} \label{eq:Mi} 
\cM_i: \qquad    \btheta_{i,t} =  \begin{pmatrix} \bgamma_{i,t}\\ \bbeta_{i,t} \end{pmatrix},  \quad
			      \F_{i,t}        =  \begin{pmatrix} \f_{i,t}\\  \bphi_t \end{pmatrix}, \qquad i=\seq1N, 
\end{equation} 
with subvectors of conformable dimensions; the  linear predictor is then $ \lambda_{i,t} = \bgamma_{i,t}'\f_{i,t} +\bbeta_{i,t}'\bphi_t$.
Here $\f_{i,t}$ contains constants and series-specific predictors-- such as item-specific prices and promotions in the sales forecasting context. 
The latent factor vector $\bphi_t$ is common to all series-- such as seasonal or brand effects in the sales forecasting context.  Each series has its own state component $\bbeta_{i,t}$ so that the impacts of   common factors   are series-specific as well as time-varying.  

In parallel, a separate model of some form-- in our case, a DLM-- also depends on $\bphi_t$ and possibly other factors. Denote this model by $\cM_0.$ Forward sequential analysis of data relevant to $\cM_0$ defines posterior distributions for $\bphi_t$ at any time $t$ that can be used to infer and forecast the $\bphi_t$ process as desired.   These inferences on the common factors are then forwarded to each model $\cM_i$ to use in forecasting the individual series. 

\subsubsection{Summary Analysis}
Consider now forecasting and then $1-$step updates and evolution from times $t-1$ to $t.$  Extend the notation for information sets $\cD_t,\cI_t$ to be model specific with model indices $i=\seq0N.$  

\medskip\noindent{\bf Forecasting:} This   is a direct  and full probabilistic extension-- enabled by simulation-- of the simple theoretical examples of \lq\lq conditioning on external forecast information'' in conditionally linear, normal models in examples of~\citealp[][(section 16.3).]{west1997book}
\begin{enumerate} \itemsep-3pt
\item \label{Mi:prior} At $t-1$ we have conditionally independent prior summaries for the series-specific states $i,$ namely $(\btheta_{i,t}| \cD_{i,t-1},\cI_{i,t-1})$ for each $i=\seq1N,$ having evolved independently from the $\btheta_{i,t-1}.$    
\item \label{M0:priorsim} Independently, model $\cM_0$ simulates the trajectory of the latent factor process into the future, i.e., generates independent samples  $\bphi_{\seq{t}{t+k}}^s \sim (\bphi_{\seq{t}{t+k}} | \cD_{0,t-1},\cI_{0,t-1})$ at time $t-1$ and for any $k\ge 0,$    where $s=\seq1S$ indexes $S$ Monte Carlo samples. 
\item \label{Mi:recouple} Send these synthetic  latent factors to each individual model. In  $\cM_i$, use $S$ parallel DCMM analyses  
to forecast over times $\seq{t}{t+k}$. Each analysis conditions on one sampled $\bphi_{\seq{t}{t+k}}^s$. One Monte Carlo draw 
from the implied predictive distribution of $y_{i,\seq{t}{t+k}} | \bphi_{\seq{t}{t+k}}^s, \cD_{i,t-1},\cI_{i,t-1}$ yields a sampled trajectory 
$y_{i,\seq{t}{t+k}}^s,$ so we create a Monte Carlo sample of size $S$ accounting for the inferences on, and uncertainty about, the latent factor process as defined under $\cM_0.$ 
\end{enumerate}   
The last step builds on the fact that  the DCMM analysis detailed earlier applies to each series-- independently across series-- conditional on a value of $ \bphi_t.$  Note also the use of parallelization.

\medskip\noindent{\bf Updating:} Observing the $y_{i,t}$ we now update in each model $\cM_i$ separately. 
\begin{enumerate} \itemsep-3pt\setcounter{enumi}{3}
\item \label{Mi:bmaprobs} 
For each $s,$ compute the value of the $1-$step ahead predictive p.d.f $p(y_{i,\seq{t}{t+k}} | \bphi_{\seq{t}{t+k}}^s, \cD_{i,t-1},\cI_{i,t-1})$  from the conditionally conjugate DGLM analysis. Use these as marginal likelihoods to evaluate implied posterior probabilities over the $s=\seq1S$ latent factor values 
relative to uniform $(1/S)$ prior probabilities. 
\item  Apply the standard DGLM updating to compute Monte Carlo sample $s-$specific posterior mean vectors and variance matrices for 
$\btheta_{i,t} | \bphi_t^s, y_{i,t}, \cD_{i,t-1},\cI_{i,t-1}.$   Marginalize over the $\bphi_t^s$ with respect to the probabilities from \ref{Mi:bmaprobs} above to deduce 
implied Monte Carlo approximations to the posterior mean vector and variance matrix of 
$\btheta_{i,t} | \cD_{i,t},$ where  $\cD_{i,t}$ now implicitly includes information from $\cD_0$ as well as $ \{ y_{i,t}, \cD_{i,t-1},\cI_{i,t-1} \}.$ 
\end{enumerate} 
Evolution to time $t+1$ now completes the cycle. 

The model  provides opportunity to explore differences across series, and over time, in effects of  latent 
factors. If the effects on series  $y_{i,t}$ are similar to those under $\cM_0,$ then $\bbeta_{i,t}$ will be close to one. Inferred trajectories of $\bbeta_{i,t}$ over time will capture relevant deviations and allow comparisons across series in how strongly they relate to the latent factors.


\section{An Example in Multi-Scale Sales Forecasting}\label{sec:multi-scale_application}

\begin{figure}[ht]
\includegraphics[width=\textwidth]{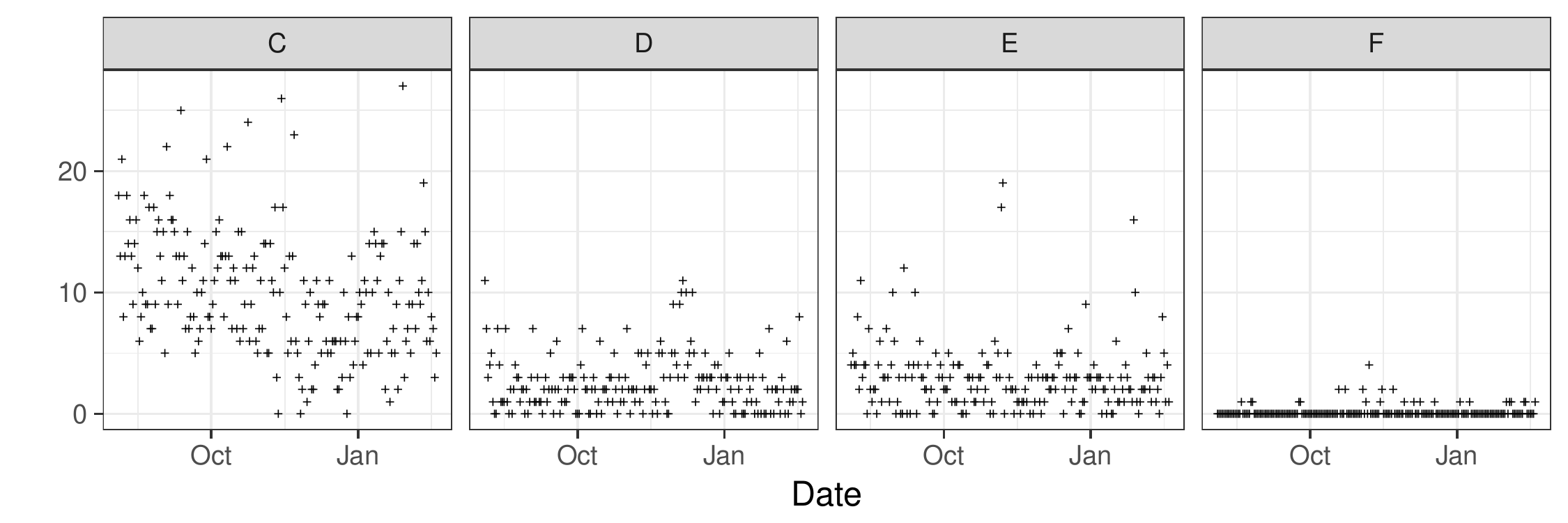}
\caption{Daily unit sales (in counts per day) of four spaghetti items C--F in one store from August 4th 2009 to February 19th 2010.}\label{fig:ms_items}
\end{figure}

\subsection{Context, Data and Models} 

Our case study involves a number of multi-scale features that offer potential for improved forecasting using the multivariate/multi-scale strategy. Items are related in product type categories, by brand, and by consumer behavior as evidenced in several ways including, simply, the day-of-week seasonality related to \lq\lq store traffic''.   We give one example here focused on this latter feature,  comparing forecasting performance of a multi-scale model-- based on one specific common pattern model $\cM_0$-- to a set of individual DCMMs.  In one selected store, we identify $N=17$ spaghetti items for the example. Figure~\ref{fig:ms_items}
 plots the daily sales of four of these items  over the period August 2009 to February 2010, giving some indication of the ranges of demand levels and patterns over time.  We take models $\cM_i$ in which the baseline DCMMs of~\eqno{Mi} have
$\f_{i,t}' =  ( 1,\log(\textrm{price}_{i,t}))$ and $\bphi_t\equiv \phi_t$ is a scalar representing \lq\lq current day-of-week effect''. 
Further, the state evolution matrix in all cases is $\G_t\equiv \G=\I$. 
We use the same DGLM specification-- but with component-specific state vectors-- for each of the binary and shifted Poisson components of these DCMMs. 

In principle,  model $\cM_0$ could be any external model used to map and predict patterns of traffic in the store impacting spaghetti purchases. 
Our example uses a simple model based on aggregate sales that highlights the relevance of the term multi-scale.  This could obviously be modified to incorporate 
additional predictors of day-of-week effects,   but serves well to illustrate the analysis here. Since pasta sells at high levels,  aggregate-level sales of types of pasta such as spaghetti are typically high and more clearly reveal the weekly demand patterns as they change day-to-day.  So a relevant aggregate series  would be a natural candidate for $\cM_0.$  We take $y_t$ to be the log of total spaghetti sales in this store on day $t$ as our example,  and specify $\cM_0$ as a flexible dynamic linear model incorporating time-varying effects and stochastic volatility as in Appendix~\ref{app:sequentiallearning}.  This DLM $\cM_0$  has a local linear trend, a regression component with the scaled log average spaghetti price as a predictor,  the full Fourier seasonal component for the $7-$day seasonal pattern over the week, and a yearly seasonal effect with the first two harmonics of  period 365.
Discount factors in this DLM were chosen based on previous analyses of aggregate sales, resulting in $\delta=0.995$ for each of the trend and regression components, $\delta=0.999$ for each of the seasonals, and $\beta=0.999$ for the residual stochastic variance process.  
Note that, at each time, the full posterior for the implied  \lq\lq current day effect'' $\phi_t$ can be trivially simulated from this model as it is simply one element of a linear transformation of the Fourier coefficients to a $7-$dimensional seasonal factor vector~\citep[][section 8.6.5]{west1997book}. In all analyses we used an initial three weeks of training data to specify initial priors for DCMM state vectors at the time index $t=1$ representing the start of the 200 day period. Analysis was then run over the first period of 100 days, prior to then forecasting  $\seq1{14}$ days ahead on each of the following 100 days.
 Predictive performance is assessed across a range of random effect discount factors
$\rho \in \{0.4, 0.6, 0.8, 1.0\}$  in each $\cM_i.$  Initial priors for the baseline and multi-scale DCMMs are matched, adjusting for
the use of the single latent seasonal factor $\phi_t$ in the latter compared to the individual Fourier models in the former. We use fixed discount factors of $\delta = 0.999$ (Bernoulli) and $0.99$ (Poisson) for each of the local level, price regression state elements, and the state elements corresponding to the seasonal effects (the vector of time-varying Fourier coefficients in the baseline DCMMs, and the single $\phi_t$ in the multi-scale model, respectively).

%

\subsection{Forecast Metrics}

We have discussed general issues of forecast evaluation and comparison in Sections~\ref{sec:eval_counts} and~\ref{sec:eval_probs}, stressing the applied need for global probabilistic metrics in general, and particularly in contexts of low count time series.   That said, for some basic comparisons of the baseline with multi-scale DCMMs,  it is also of interest to relate to traditional point forecast accuracy measures. We do that here with three empirical loss functions from the count forecasting 
literature~\citep[e.g.][]{fildes2007against},  namely scaled mean squared error (sMSE), mean absolute deviation (MAD), and mean ranked probability score (MRPS) metrics.
Metrics are specific to a chosen lead-time $k>0.$  
For any series $y_t$,  denote by $f_{t,k}$ a point forecast of $y_{t+k}$  made at time $t$.  
The scaled squared error (sSE) based on outcome $y_{t+k}$ is   $sSE_t(k) = (y_{t+k}-  f_{t,t+k})^2/(\bar y_t)^2$ where $\bar y_t$ is the mean of $y_{\seq1t}.$ 
Then the sMSE for forecast horizon $k$ is calculated as the average of the $sSE_t(k)$ over all days $t.$ This has become of interest in evaluating point forecast accuracy of count data as it is well defined (unless all observed values are zero), and it is not as sensitive to high forecast errors as the 
MSE~\citep{kolassa2016evaluating}.  The optimal point forecast under sMSE is the $k-$step ahead predictive mean.
The MAD metric is standard, simply the time average of $|y_{t+k}- f_{t,t+k}|,$ and the optimal point forecast is the $k-$step ahead predictive median. 
The third metric RPS is a scoring rule related to earlier discussed and utilized PIT measures~\citep{snyder2012forecasting, kolassa2016evaluating}. If the time $t$ forecast distribution for $y_{t+k}$ has c.d.f. $P_{t,k}(\cdot),$ then  $RPS_t(k) = \sum_{j = 0}^{\infty} (P_{t,k}(j) - \ind{y_{t+k} \leq j})^2,$ and we calculate the MRPS for forecast horizon $k$ by averaging $RPS_t(k)$ across all days $t$.

\subsection{Forecasting Analysis and Comparisons}

Metrics from the analyses are shown  in Figure~\ref{fig:ms_items}. These items represent the different levels of demand present in this store: high demand (item C), moderate demand (items D, E), and low demand (item F). For each of the three metrics, we  evaluate  forecasts across $\seq1{14}$ days ahead at each day. The baseline and multi-scale models are evaluated across a range of random effects discount factors $\rho \in \{0.4, 0.6, 0.8, 1.0\}$. Since different values of $\rho$ may provide better forecasts across the forecasting horizon,  we display only the minimum results across each of the four models. That is, for illustration here we only present the results from the best baseline and multi-scale model for each item, error metric, and forecast horizon.

\begin{figure}[htbp!]
\begin{center}
\includegraphics[width=\textwidth]{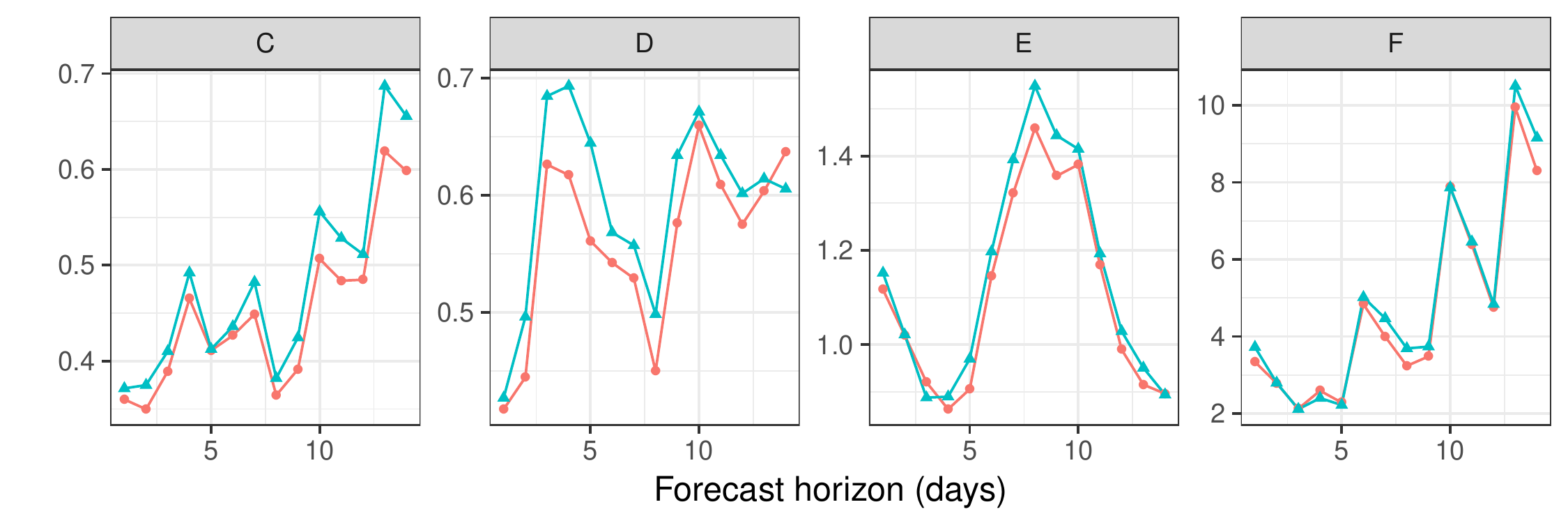}
\caption{Scaled mean squared error (sMSE) vs forecast horizon (days) for items C, D, E, and F from the multi-scale (orange circles) and baseline (blue triangles) models.}\label{fig:ms_sSE}
\bigskip
\includegraphics[width=\textwidth]{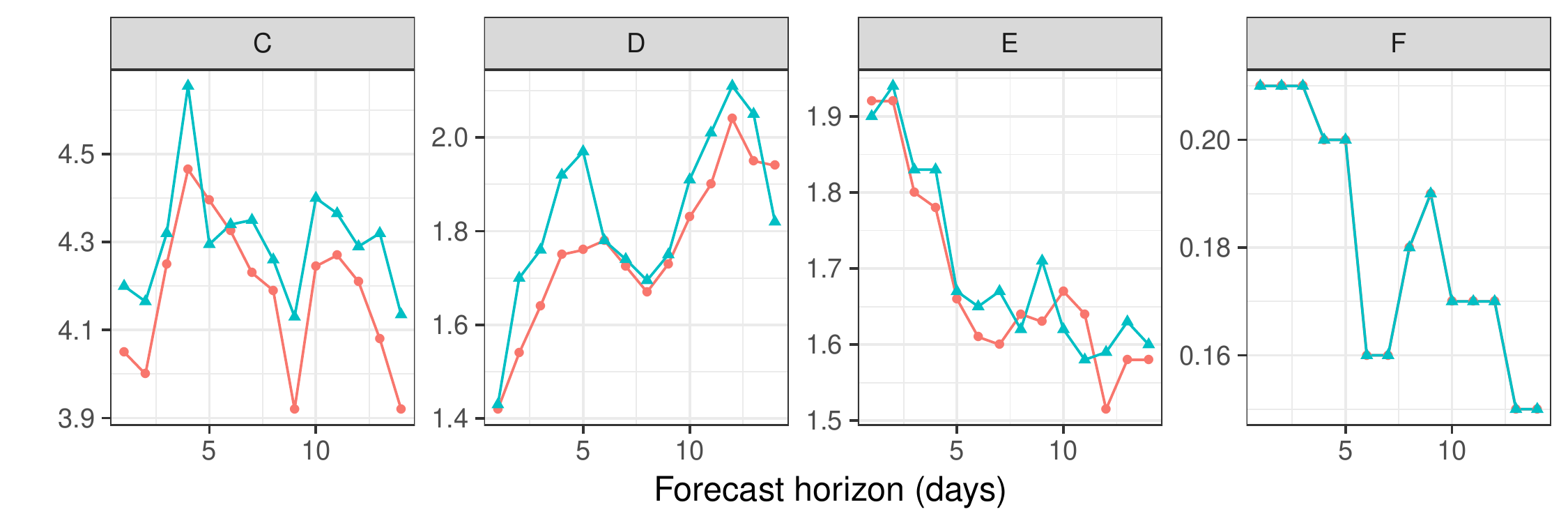}
\caption{Mean absolute deviation (MAD) vs forecast horizon (days) for items C, D, E, and F from the multi-scale (orange circles) and baseline 
(blue triangles) models. }\label{fig:ms_mad}
\bigskip
\includegraphics[width=\textwidth]{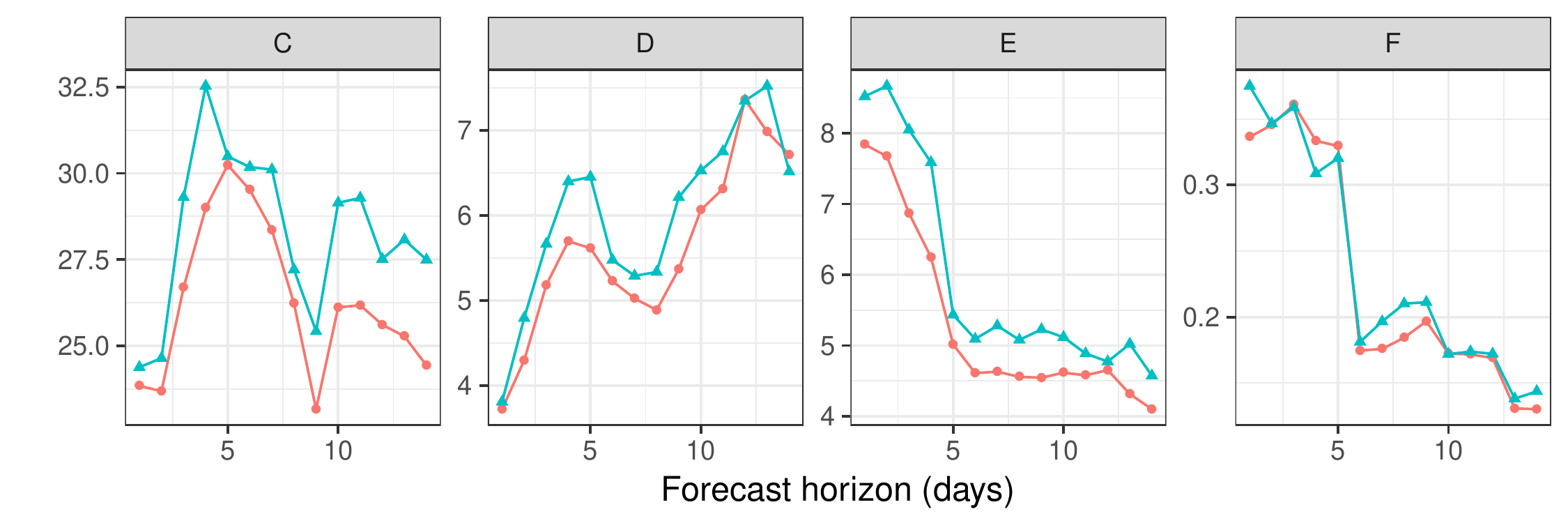}
\caption{Mean rank probability score (MRPS) vs forecast horizon (days) for items C, D, E, and F from the multi-scale (orange circles) and baseline (blue triangles) 
models.}\label{fig:ms_rps}
\end{center}
\end{figure}

\paragraph{Comparisons under sMSE:} 
Figure~\ref{fig:ms_sSE} shows the sMSE versus forecast horizon  for each item from the best performing multi-scale and baseline DCMMs.  Comments by specific items are as follows. 
\begin{itemize}\itemsep=-3pt
\item[C:] The multi-scale DCMM has lower sMSE across the entire forecasting horizon. The sMSE of both models increases as the forecasting horizon gets longer, however the percentage decrease in sMSE of the multi-scale vs baseline DCMM generally increases for longer forecast horizons. The greatest percentage decreases in sMSE of $8-10\%$ occur when forecasting 10, 11, 13, and 14 days ahead. For 10 out of the 14 forecast horizons, we see at least a 5\% decrease in sMSE. Averaging the sMSE across the forecasting horizon, the overall percentage decrease of the multi-scale DCMM versus the baseline DCMM is 6\%. 
\item[D:] The multi-scale DCMM has lower sMSE than the baseline DCMM when forecasting 1 to 13 days ahead. For this item, the percentage decrease in sMSE of the multi-scale method is larger for short- to mid-term forecasting. When forecasting 2, 4, 5, 8, and 9 days ahead, the percentage decrease in sMSE is between $9-13\%.$ When forecasting 10-13 days ahead, the percentage decrease in sMSE drops to between $1-5\%.$ When forecasting 14 days ahead, the multi-scale DCMM has a 5\% decrease in sMSE compared to the baseline DCMM. However, averaging the sMSE across the forecasting horizons, the average percent decrease in sMSE of the multi-scale DCMM versus the baseline DCMM is 6\%. 
\item[E:] The multi-scale DCMM has lower sMSE for 12 out of the 14 forecast horizons. The largest percentage decreases in sMSE of $4-7\%$ occur for mid range forecasting ($5-9$ steps ahead). Averaging the sMSE across the forecasting horizon, the multi-scale DCMM decreases sMSE by about 3\% compared to the baseline DCMM. 
\item[F:] The multi-scale DCMM has lower sMSE for 10 out of the 14 forecast horizons. The largest percent decreases in sMSE of $9-13\% $ occur randomly through the forecasting horizon at 1, 7, 8, and 14 days ahead. Averaging sMSE across the forecasting horizon, we see an overall percent decrease of about 4\% with the multi-scale DCMM. 
\end{itemize}

\paragraph{Comparisons under MAD:} 
Figure~\ref{fig:ms_mad} shows the mean absolute deviations (MAD) versus forecasting horizon for each item from the best performing multi-scale and baseline DCMMs. Item specific differences are now noted. 
\begin{itemize}\itemsep=-3pt
\item[C:] The multi-scale DCMM has lower MAD than the baseline DCMM for 13 of 14 forecasting horizons. 
When forecasting 13, 14 days ahead, the MAD of multi-scale DCMM is 5-6\% lower than the baseline DCMM. Averaging across the forecasting horizon, the MAD of the multi-scale DCMM is 2.8\% less than the MAD of the baseline DCMM. 
\item[D:] The multi-scale DCMM has lower MAD for 13 of 14 forecasting horizons. The multi-scale approach has the greatest improvement in MAD of 6-11\% when forecasting 2 to 5 days ahead. Across the entire forecasting horizon, the multi-scale DCMM decreases MAD by about 4\% compared to the baseline DCMM. 
\item[E:] The multi-scale DCMM has lower MAD for 10 of 14 forecasting horizons. The difference in MAD between these two approaches seems minimal, although there is a 1.2\% decrease in MAD across the forecasting horizon. 
\item[F:] the multi-scale and standard DCMMs have the exact same MAD across the forecasting horizon. Referring back to Figure~\ref{fig:ms_items}, we see that item F has zero sales on more than 50\% of days. Each DCMM predicts a zero response with at least 50\% probability, making the median forecast zero over time. Since each method has the same median forecast, the MAD is the same at each time point. 
\end{itemize}

\paragraph{Comparisons under RPS:} 
Figure~\ref{fig:ms_rps} shows the ranked probability score (RPS) versus the forecasting horizon for each item from the best performing multi-scale and baseline DCMMs. Item specific differences are as follows. 
\begin{itemize}\itemsep=-3pt
\item[C:] The multi-scale DCMMs have lower RPS than the baseline DCMMs for 14 of 14 forecasting horizons. When forecasting 9 to 14 days ahead, the RPS of the multi-scale methods is about $7-11\%$ lower than the standard methods. Averaged across the forecasting horizon, the multi-scale models have a 7\% decrease in RPS. 
\item[D:] The multi-scale DCMMs have lower RPS for 12 of 14 forecasting horizons. When forecasting 2, 4, 5, and 9 days ahead, the percent decrease in RPS is greater than 10\%. Across the entire forecasting horizon, the percentage decrease in RPS using the multi-scale method is about 7\%. 
\item [E:] The multi-scale DCMMs have lower RPS for all 14 forecasting horizons. We see that the RPS actually decreases as the forecasting horizon gets longer, indicating that the performance of the predictive intervals may be improving for longer term forecasts. For 8 of the 14 forecasting horizons, we see percent decreases of at least 10\% in the RPS, including a 17\% decrease when forecasting 4 days ahead. Across the entire forecasting horizon, the percent decrease in RPS under the multi-scale method is about 11\%. 
\item[F:] the multi-scale DCMMs have lower RPS for 10 of 14 forecasting horizons. Across all forecasting horizons, the percentage decrease in RPS under the multi-scale method is about 3\% compared to the baseline DCMMs. 
\end{itemize} 

%
%


\section{Summary Comments}
\label{sec:conc}
 
In the context of a motivating case study and application in consumer sales and demand forecasting, we have introduced a novel class of dynamic state-space models for time series of non-negative counts, and a formal multivariate extension for many related series.
The univariate DCMM framework builds on and extends prior approaches to univariate count time series, contributing a flexible, customizable class of models. The ability to explore and include covariates as potential predictors of both binary and positive count series is of interest in many areas, and the coupling of DGLMs for these two components addresses a very common need with count data. Motivated by problems in consumer demand and sales forecasting, the opportunity to apply these models and extend their use in commercial and socio-scientific forecasting is evident. In addition to the dynamic regression components, the time-varying state-space framework allows evaluation of changes over time in regressions, trends, seasonal patterns and other forms of predictor information, and adaptability to any such changes. The Bayesian framework defines probabilistic forecasts that, accessed trivially computationally through direct, forward simulation of predictive distributions, enables evaluation of various summary measures of uncertainty about forecast paths of time series into the future and arbitrary functions of sets of future outcomes. This is important in applications as a general matter of properly communicating forecast information, and also provides the basis for formal decision analysis in decision contexts reliant on forecasts. Our examples developed in consumer sales forecasting highlight the machinery of model fitting and forecasting, and demonstrate the utility of DCMMs with series exhibiting quite differing patterns and levels of outcome intensity. This is important in applications involving many series where it is desirable to have a single model class that is flexible and adaptable to individual series characteristics. A number of traditional point forecast metrics are also discussed, along with the point that they should always be considered so long as the background context supports the role of the implicit loss function underlying any specific point forecast. More broadly on forecast assessment and evaluation, we have stressed and exemplified the use of analyses addressing the full forecast distributions, to include frequency calibration of both binary and positive count models, empirical coverage of nominal forecast intervals.
 
The embedding of sets of DCMMs into a multivariate system defines a novel class of state-space models for many related time series of
counts. Importantly, this maintains the flexibility of modeling at the univariate series level, using individual DCMMs that are linked across series via common latent factors. The linkages are series-specific, potentially time-varying random effects, so defining an overall, flexible hierarchical dynamic model framework.
Also, critically, the new multivariate/multi-scale approach maintains the ability to run fast, sequential Bayesian analysis of decoupled univariate analyses of many series, with recoupling across series based on information about common factors flowing from a parallel external model. This strategy enables analytic computations and trivial forward simulation for sequential analysis and forecasting, and by design is scalable to many series (computations grow only linear with the number of series). Our example in the motivating consumer sales case study involves a common factor process related to shared seasonal patterns and in which the external model generating inferences on the factor process is a dynamic model applied to aggregate data in which the pattern is more precisely identified. In future applications, the shared latent factor process will be multivariate, with dimensions reflecting different ways in which series are conceptually related. In product demand forecasting, for example, products can be grouped by product family, brand, store location and other factors, and both aggregate-level and external economic or business models may provide inputs to forecast several common factors representing the relevant cross-series linkages. Our pilot example illustrates the ability of the multivariate/multi-scale approach to improve forecasts at the individual series level, in both short and longer-term forecasting, and across series with intermittent, moderate, and high demand patterns. While forecast accuracy improvements cannot be expected for all series all of the time, even small increases in forecast accuracy on a number of items can have a profound impact on retail decision-making and costs. Further studies to explore the models across a very large number of series will improve understanding of pros-and-cons of the multivariate approach across a diverse range of applications. One important factor that plays a role here is the length of historical data available for each item. It is to be expected that series with shorter histories will most immediately benefit from the multivariate approach, as common features impacting demand on similar items will feed information relevant to forecasting the newer items, a context of clear interest in commercial settings when new or modified products are introduced.

\newpage

\appendix
\section{Appendix: Technical Details of DGLMs and DCMMs  \label{app:dglm} }

\subsection{Sequential Learning in DCCMs from DGLM Components \label{app:sequentiallearning} } 

Consider a DGLM in the general class as discussed and referenced in Section~\ref{sec:basicDGLM}. The  observation model has
$p(y_t \mid \eta_t, \phi) = b(y_t, \phi) \exp{ [\phi \{y_t \eta_t - a(\eta_t)\} ]} $
with natural parameter $\eta_t$ and linear predictor $\lambda_t = g(\eta_t)$ based on link function $g(\cdot).$ 
Also,  as in  \eqno{DGLMregnevo},  $\lambda_t = \F'_t \btheta_t$ where 
$\btheta_t = \G_t \btheta_{t-1} + \bomega_t$  with  $\bomega_t \sim (\bzero, \W_t)$, the latter denoting zero mean vector and  variance matrix $\W_t.$ 
The standard DGLM analysis~(\citealp{west1985dynamic};~\citealp{west1997book}~chapter~15;~\citealp{Prado2010}~section~4.4)
has the following features. 
\begin{enumerate} \itemsep=-3pt
\item  At any time $t-1$, the current information is summarized via the mean vector and variance matrix of the posterior for the current state vector, namely 
  $(\btheta_{t-1} \mid \cD_{t-1}, \cI_{t-1}) \sim [\m_{t-1}, \C_{t-1}]$. 
 \item \label{dglm:Prior} Through the evolution equation this induces $1-$step ahead prior moments on the state vector of the form $(\btheta_{t} \mid \cD_{t-1}, \cI_{t-1}) \sim [\a_{t}, \R_{t}]$ with $\a_t = \G_{t} \m_{t-1}$ and $\R_{t} = \G_{t} \C_{t-1} \G'_t + \W_t$. 
 \item The variational Bayes concept then applies to choose a conjugate prior for $\eta_t$, denoted by $(\eta_t \mid \cD_{t-1}, \cI_{t-1}) \sim \text{CP}(\alpha_t, \beta_t)$ with form
$
p(\eta_t \mid \cD_{t-1}, \cI_{t-1}) = c(\alpha_t, \beta_t) \exp{\{ \alpha_t \eta_t - \beta_t a(\eta_t)\}}.
$
Here $c(\cdot,\cdot)$ is a function of the hyper-parameters of known form depending on the specific exponential family model.  

\item \label{dglm:VBprior} The hyper-parameters $\alpha_t$ and $\beta_t$ are evaluated so that the conjugate prior satisfies the prior moment constraints 
$$\text{E}[\lambda_t \mid \cD_{t-1}, \cI_{t-1}]  = f_t =\F'_t \a_t \quad\textrm{and}\quad   \text{V}[ \lambda_t  \mid \cD_{t-1}, \cI_{t-1} ]  = q_t = \F'_t \R_t \F_t.$$
\item \label{eq:expfamforecast} Forecasting $y_t$  $1-$step ahead uses the conjugacy-induced predictive distribution with p.d.f.   
$p(y_t \mid \cD_{t-1},\cI_{t-1}) = b(y_t, \phi) c(\alpha_t, \beta_t)/c(\alpha_t + \phi y_t, \beta_t + \phi).$ 

\item  On observing $y_t,$ the posterior for $\eta_t$ has the conjugate  form of  $(\eta_t \mid \cD_t) \sim CP(\alpha_t + \phi y_t, \beta_t + \phi).$ 
\item  \label{dglm:VBpost} Under this posterior,  mapping back to the linear predictor $\lambda_t = g(\eta_t)$  implies posterior mean and variance 
$g_t = \text{E}[\lambda_t \mid \cD_t ] $ and $p_t = \text{V}[\lambda_t \mid \cD_t ].$ 
\item \label{dglm:VBpostmoms}  Finally, linear Bayes updating implies  posterior mean vector and variance matrix of the state vector as
$ (\btheta_t \mid \cD_t ) \sim [\m_t, \C_t]$ given by
$$  \m_t = \a_t +  \R_t \F_t (g_t - f_t) /q_t\quad\textrm{and}\quad  \C_t = \R_t - \R_t \F_t \F'_t \R'_t (1 - p_t/q_t)/q_t.$$
This completes the time $t-1$-to-$t$ evolve-predict-update cycle.
\end{enumerate} 
Some of the structure and computations implied require comment and are highlighted in the key cases of interest for count data.  In each case, the link function $g(\cdot)$ is the identity so that $\eta_t=\lambda_t.$ 

\medskip\noindent{\bf\em Bernoulli logistic DGLM: }  Here the series $y_t$ is relabelled as $z_t = 0/1$ with
$z_t \sim Ber(\pi_t)$ and $ \eta_t = \text{logit}(\pi_t)$.   In the exponential family p.d.f. form the terms are $\phi=1$, $b(y_t,\phi) =1$ and  $a(\eta_t) = \log(1+\exp(\eta_t)).$ 

The conjugate prior in part~\ref{dglm:VBprior} above is Beta, $\pi_t \sim Be(\alpha_t, \beta_t)$, with the hyper-parameters defining  $f_t = \psi(\alpha_t) - \psi(\beta_t)$ and $q_t = \psi'(\alpha_t) + \psi'(\beta_t)$, where $\psi(\cdot)$ and $\psi'(\cdot)$ are the digamma and trigamma functions, respectively. The values $(\alpha_t,\beta_t)$ can be trivially computed from $(f_t,q_t)$ via iterative numerical solution based on standard Newton-Raphson.  The $1-$step ahead forecast  is Beta-Bernoulli with   $(z_t \mid \cD_{t-1},\cI_{t-1}) \sim BBer(1, \alpha_t, \beta_t)$ defined simply by 
$\text{Pr}[z_t=1| \cD_{t-1},\cI_{t-1}]= \alpha_t/(\alpha_t+\beta_t).$     The updated moments of the linear predictor in part~\ref{dglm:VBpost} above are then  trivially computed via the equations $g_t = \psi(\alpha_t+z_t) - \psi(\beta_t+1-z_t)$ and $p_t = \psi'(\alpha_t+z_t) + \psi'(\beta_t+1-z_t).$

\medskip\noindent{\bf\em Poisson loglinear DGLM:}  Here  $y_t \sim Po(\mu_t)$ with $ \eta_t = \log(\mu_t)$.    In the exponential family p.d.f. form the terms are $\phi=1$, $b(y_t,\phi) =1/y_t!$ and  
$a(\eta_t) = \exp(\eta_t).$  

The conjugate prior
in part~\ref{dglm:VBprior} above is Gamma, $\mu_t \sim Ga(\alpha_t, \beta_t)$, with the hyper-parameters defining  $f_t = \psi(\alpha_t) - \log(\beta_t)$ and $q_t = \psi'(\alpha_t).$  The values $(\alpha_t,\beta_t)$ can be trivially computed from $(f_t,q_t)$ via iterative numerical solution based on standard Newton-Raphson.  The $1-$step ahead forecast  is negative binomial, $(y_t \mid \cD_{t-1},\cI_{t-1}) \sim Nb(\alpha_t, \beta_t/(1 + \beta_t)$.  The updated moments of the linear predictor in part~\ref{dglm:VBpost} above are  trivially computed via the equations $g_t = \psi(\alpha_t+y_t) - \log(\beta_t+1)$ and $p_t = \psi'(\alpha_t+y_t).$

\medskip\noindent{\bf\em Normal DLM:} We also note the special case of normal models when the DGLM reduces to a conditionally normal DLM.   This is of relevance to count time series in case of large counts where a log transform-- for example-- of the count series can often be well-modeled using a normal DLM as an approximation. This also allows for inclusion of volatility via a time-varying conditional variance.    With   $y_t$ the logged values of the original count series, a normal model has
 $y_t \sim N(\mu_t,v_t)$ with $ \eta_t=  \mu_t$.    In the exponential family p.d.f. form the term $\phi$ becomes, generally, a time-dependent precision,
  $\phi_t=1/v_t$,  while $b(y_t,\phi_t) =(\phi_t/2\pi)^{1/2} \exp(-\phi_ty_t^2/2)$ and  
$a(\eta_t) = \eta_t^2/2.$ 

The conjugate prior
in part~\ref{dglm:VBprior} above is normal, $\mu_t \sim N( a_t, A_t v_t) $ which matches the general conjugate form with 
$\alpha_t = a_t/A_t$ and $\beta_t = 1/A_t.$   Prior to posterior updating in part~\ref{dglm:VBpostmoms} reduces to a standard Kalman filter update.  When embedded in the DLM, the   additional  assumption that the evolution noise terms $\bomega_t$ in \eqno{DGLMregnevo}
are also normal implies that DGLM evolution/updating equations are exact in this special case.  However,  for most practical applications it is relevant to also estimate the conditional variances  $v_t = 1/\phi_t.$ The simplest and most widely-used extension is that based on a standard Beta-Gamma stochastic volatility model for $\phi_t$ which,  is analytically tractable.  The resulting theory is then based on normal/inverse gamma prior and posterior distributions for $(\mu_t,v_t). $ Details of the resulting modifications to forward filtering and forecasting analysis are very standard~(\citealp{west1997book}~chapter~4 and section~10.8;~\citealp{Prado2010}~section~4.3).

\subsection{Discount Factor Model Specifications \label{app:discountfactors} } 

\subsubsection{Traditional Component Discounting} 
Specification of the required evolution variance matrices $\W_t$ in  \eqno{DGLMregnevo} uses the standard, parsimonious and effective discount method based on component discounting~\citep[][chapter 6]{west1997book}. In most practical models the state vector is naturally partioned into components representing different 
explanatory effects, such as trends (e.g., local level, local gradient), seasonality (time-varying seasonal factors or Fourier coefficients) and effects of independent predictor variables. That is, for some integer $J$ we have $\btheta_t' = (\btheta_{t1}',\ldots,\btheta_{tJ}')$. It is natural to define $\W_t$ to represent potentially differing degrees of stochastic variation in these components and this is enabled using separate discount factors $\delta_1,\ldots,\delta_J,$ where each $\delta_j\in (0,1].$ A high discount factor implies a low level of stochastic change in the corresponding elements of the state vector, and vice-versa (with $\delta_j=1$ implying no stochastic noise at all-- obviously desirable but rarely practically relevant).   The definition of $\W_t$ is as follows.  

From Appendix~\ref{app:sequentiallearning} part \ref{dglm:Prior} above, the time $t-1$ prior variance matrix of $ \G_t\btheta_{t-1}$  is $\P_{t} = \G_{t} \C_{t-1} \G'_t;$  this represents information levels about the state vector following the deterministic evolution via $\G_t$ but before the impact of the  evolution noise that then simply adds $\W_t.$   Write 
$\P_{tj}$ for the diagonal block of $\P_t$ corresponding to state subvector $\btheta_{tj}$ and set 
$$\W_t = \text{block diag}[ \P_{t1}(1-\delta_1)/\delta_1 , \ldots, \P_{tJ}(1-\delta_J)/\delta_J].$$ Then the implied prior variance matrix of $\btheta_t$ following the evolution has corresponding diagonal block elements $\R_{tj} = \P_{tj}/\delta_j$ while maintaining off-diagonal blocks from $\P_t.$ 
Thus, the stochastic part of the evolution increases uncertainties about state vector elements in each subvector $j$ by $100(1-\delta_j)/\delta_j\%,$ 
maintains the correlations in $\P_{tj}$ for state elements within the subvector $j,$ while reduces cross-correlations between state vector elements in differing subvectors. In practice, high values of the $\delta_j$ are desirable and typical applications use values in the range $0.97-0.99$ with, generally, robustness in terms of forecasting performance with respect to values in the range. Evaluation of forecast metrics on training data using different choices of discount factors is a basic strategy in model building and tuning.

\subsubsection{Discount Factor Specifications for Random Effects Extensions of DGLMs \label{app:discountrandomeffects} } 

We use the random effects extension of Section~\ref{sec:randomeffect} for the shifted Poisson case in evaluating forecasts of non-zero count series. 
As detailed in that section, this is enabled by extension of the state vector to include time-specific zero mean elements defining these random effects. 
Practically, this is defined using a random effects discount factor $\rho\in (0,1]$ whose net impact 
on the DGLM analysis summarized in Appendix~\ref{app:sequentiallearning} is simply to inflate the prior variance of the linear predictor
$\lambda_t = \log(\mu_t).$   That is, $q_t = \text{V}[ \lambda_t  \mid \cD_{t-1}, \cI_{t-1} ] $ in
Appendix~\ref{app:sequentiallearning} part~\ref{dglm:VBprior}
is modified to  $q_t+v_t$ where  $v_t =q_t  (1-\rho)/\rho$, resulting in $q_t/\rho.$  As with the standard discounting on state vectors above, 
evaluation of forecast metrics on training data using different choices of $\rho$ is a basic strategy for choosing values, and these will be specific to each time series. 
   
More theoretical insight can be gained by considering the impact on the implied $1-$step forecast distributions. 
In the standard DGLM with no random effects, recall that the conditional prior for the Poisson mean $\mu_t$ is $Ga(\alpha_t,\beta_t)$ where the parameters are chosen to be consistent with the prior mean $f_t$ and variance $q_t$ of $\lambda_t = \log(\mu_t),$ namely $f_t = \psi(\alpha_t) - \log(\beta_t)$ and $q_t = \psi'(\alpha_t).$   Suppose we have a relatively precise prior-- with $q_t$ modestly high-- so that the approximations 
$\psi(\alpha_t) \approx \log(\alpha_t)$ and $\psi'(\alpha_t) \approx 1/\alpha_t$ are valid (these are in fact very accurate approximations in many applications). 
Then $f_t \approx \log(\alpha_t/\beta_t)$ and $q_t \approx 1/\alpha_t,$ resulting in $\alpha_t = 1/q_t$ and $\beta_t  = \alpha_t \exp(-f_t).$ 
The implied $1-$step forecast distribution for $y_t$ is negative binomial with mean $\alpha_t/\beta_t = \exp(f_t)$ and variance 
 $\alpha_t/\beta_t + \alpha_t/\beta_t^2 = \exp(f_t) ( 1 + \exp(f_t) q_t).$ 
 
 Now consider the impact of the random effects model extension.  As noted above, the practical impact of the discount factor $\rho$ is that $q_t$ is inflated to 
 $q_t/\rho.$ The resulting negative binomial forecast distribution then has the same mean $\exp(f_t)$-- not impacted at all by $\rho$-- but now has   variance 
 $\exp(f_t) ( 1 + \exp(f_t) q_t/\rho ).$    This has the same base component $\exp(f_t)$ (the \lq\lq Poisson'' component) but  the second term (considered the \lq\lq extra-Poisson'' variation in the negative binomial) increases by a factor of $1/\rho$.  Note that the impact of the random effects extension is then to increase 
 forecast variances more at higher levels of the series (higher $f_t$), consistent with the aim of improving forecasts for infrequent higher events.

\subsection{Forecasting in DCMMs \label{app:forecasting}} 

The compositional nature of the DCMM yields access to a full predictive distribution at a future time point that is a mixture of the forecast distributions implied for each of the independent Bernoulli and shifted Poisson DGLMs. Forecasting $k-$steps ahead from time $t$,   the forward evolution of DGLM state vectors over times $\seq{t+1}{t+k}$ and variational Bayes' constraint to conjugate forms for implied Bernoulli and Poisson parameters yields analytic tractability and trivial computation. 
That is, at time $t$, the $k-$step ahead forecast distribution has a p.d.f. of the compositional form 
$$
p(y_{t+k}\mid \cD_t,\cI_t, \pi_{t+k}) = (1- \pi_{t+k}) \delta_0(y_{t+k}) + \pi_{t+k} h_{t,t+k}(y_{t+k})$$
where: 
\begin{itemize} \itemsep=-3pt
\item  $(\pi_{t+k} \mid \cD_t,\cI_t) \sim Be(\alpha_t^0(k), \beta_t^0(k))$ and $\delta_0(y)$ is the Dirac delta function at zero. 
\item $h_{t,t+k}(y_{t+k})$ is the density function of $y_{t+k}=1+x_{t+k}$ where $x_{t+k}$ has the negative binomial distribution 
$$ (x_{t+k}\mid \cD_t,\cI_t) \sim Nb \Big( \alpha^+_t (k), \frac{\beta_0^+(k)}{1 + \beta_t^+(k)} \Big). $$
\item  The defining parameters $\alpha_t^0(k)$, $\beta_t^0(k)$, $\alpha_t^+(k)$, $\beta_t^+(k)$   are computed from the binary and positive count DGLMs, respectively.
\end{itemize} 
That is, the mixture places probability $1- \pi_{t+k}$ on $y_{t+k}=0$, and probability  $\pi_{t+k}$ on the implied shifted negative binomial distribution. 

Depending on the forecasting context, we may be interested in marginal or joint forecasting. The above details define the relevant marginal forecast distributions at any time.
However, we are generally much more interested joint forecasts for $y_{t+1}, \ldots, y_{t+k}$ and, as discussed in Section~\ref{sec:dcmmmodel}, in forecasting paths with an opportunity to explore dependencies in outcomes  over time as well as to predict functions of them.   This is trivially-- and computationally efficiently-- enabled using forward simulation, as follows. 
\begin{itemize} \itemsep=-3pt
\item  At time $t$,   propagate the posterior $(\btheta_t \mid \cD_t)$ to the $1-$step ahead prior $(\btheta_{t+1} \mid \cD_t,\cI_t)$. 
\item Apply the variational Bayes' constraint to the implied conjugate prior on $\eta_{t+1}$,  so that the implied $1-$step ahead forecast distribution is of the form
given in part~\ref{eq:expfamforecast} of Appendix~\ref{app:sequentiallearning}; that is, Beta-Bernoulli or negative binomial depending on the chosen DGLM. 
\item Simulate an outcome $y_{t+1}^*$ from this $1-$step forecast distribution. 
\item Treating this synthetic outcome as \lq\lq data'',   perform the time $t+1$ update step to revise the prior to posterior for the state vector, now based on 
modified information $\cD_{t+1}$ in which the unknown $y_{t+1}$ is substituted by its synthetic value $y_{t+1}^*.$ 
\item Evolve to time $t+2$ and repeat the process to simulate a synthetic $y_{t+2}^*$ and then update based the model. 
\item Continue this process over lead-times $t+3,\ldots, t+k$ to the chosen forecast horizon. 
\end{itemize} 
This results in one synthetic path $y_{t+1}^*, \ldots, y_{t+k}^*$ defining  a single Monte Carlo sample from the joint distribution of $y_{t+1}, \ldots, y_{t+k}$
conditional on $\{  \cD_t,\cI_t \}.$  Repeat this process many times to generate a large Monte Carlo sample from this joint predictive distribution.  At each marginal time point $t+j,$ the corresponding samples give a Monte Carlo representation of the predictive distribution at that lead-time, while the full sample provides the required opportunities for inference on paths, dependencies of outcomes between time points, and functions (cumulative outcomes, exceedance over some specific level, etc) of the path into the future.

%

\end{document}